\pdfoutput=1
\documentclass[11pt]{article}

\usepackage[final]{acl}
\usepackage{pgfplots}
\pgfplotsset{compat=1.18} % 或其他版本号，确保兼容性
\usepackage{times}
\usepackage{latexsym}
\usepackage[table,xcdraw]{xcolor}
\usepackage{multirow}
\usepackage{graphicx}
\usepackage{adjustbox}
\usepackage{wrapfig}

\usetikzlibrary{positioning,matrix,calc,arrows.meta,decorations.pathreplacing,shapes.geometric}
\definecolor{mygreen}{RGB}{46,139,87}
\definecolor{myred}{RGB}{255,152,150}
\definecolor{myblue}{RGB}{30,144,255}
\definecolor{myyellow}{RGB}{219,219,141}
\definecolor{mybrown}{RGB}{197,157,148}
\usepackage{pgfplotstable}

\usepackage{xspace}

\newcommand{\our}{\textsc{ChainFed}\xspace}

\usepackage{microtype}
\usepackage{booktabs} % for professional tables
\usepackage{multirow}
\usepackage[table,xcdraw]{xcolor}

\usepackage{amsmath}
\usepackage{amssymb}
\usepackage{mathtools}
\usepackage{tikz}
\usepackage{pgfplots}

\usepackage{algorithm}
\usepackage{algpseudocode}
\usepackage{subcaption}
\usepackage{graphicx} 
\definecolor{grayrow}{gray}{0.9} % 定义最后一行背景色
\definecolor{algo_col}{HTML}{b0dedb}

\usepackage{arydshln}
\usepackage{pifont}
\usepackage{colortbl}  
\usepackage{booktabs}
\usepackage[T1]{fontenc}
\usepackage[utf8]{inputenc}

\usepackage{microtype}

\usepackage{inconsolata}

\usepackage{graphicx}

\title{Beyond End-to-End: Dynamic Chain Optimization for Private LLM Adaptation on the Edge}

\author{
Yebo Wu$^{1}$\footnotemark[2], Jingguang Li$^{1}$\footnotemark[2], Chunlin Tian$^{1}$, Kahou Tam$^{1}$, Zhijiang Guo$^{2}$\footnotemark[1], Li Li$^{1}$\footnotemark[1] \\
$^1$State Key Laboratory of IOTSC, University of Macau \\
$^2$Hong Kong University of Science and Technology (Guangzhou) \\
\texttt{\{yc37926, mc45005, yc27402, yc37436, llili\}@um.edu.mo} \\
\texttt{zhijiangguo@hkust-gz.edu.cn} \\
}

\begin{document}
\maketitle
\renewcommand{\thefootnote}{\fnsymbol{footnote}}
\footnotetext[1]{Corresponding authors.}
\footnotetext[2]{Equal contribution.}
\renewcommand{\thefootnote}{\arabic{footnote}}

% \begin{document}
% \maketitle

\begin{abstract}

Federated fine-tuning enables privacy-preserving LLM adaptation but faces a critical bottleneck: the disparity between LLMs' high memory demands and edge devices' limited capacity.
To break the memory barrier, we propose \textbf{Chain Federated Fine-Tuning (\our)}, an innovative paradigm that forgoes end-to-end updates in favor of a sequential, layer-by-layer manner. It first trains the initial adapter to convergence, freezes its weights, and then proceeds to the next. This iterative train-and-freeze process forms an optimization chain, gradually enhancing the model's task-specific proficiency.
\our further integrates three core techniques: 1) Dynamic Layer Co-Tuning to bridge semantic gaps between sequentially tuned layers and facilitate information flow; 2) Globally Perceptive Optimization to endow each adapter with foresight beyond its local objective; 3) Function-Oriented Adaptive Tuning to automatically identify the optimal fine-tuning starting point.
Extensive experiments on multiple benchmarks demonstrate the superiority of \our over existing methods, boosting average accuracy by up to 46.46\%.

% On text classification tasks, \our boosts average accuracy by up to 46.46\%, while on instruction tuning tasks, it improves average performance by 6.09\% and simultaneously reduces peak memory usage by 4.60$\times$.

\end{abstract}

\section{Introduction}
\label{sec:intro}

% Large Language Models (LLMs)~\cite{liu2024deepseek,bai2023qwen} are revolutionizing mobile intelligence, but unlocking their full potential requires further fine-tuning them for specific downstream tasks. This adaptation, however, is often hindered by the fact that task-specific data is siloed on user devices, shielded by privacy regulations that make centralized training impractical~\cite{mcmahan2017communication}.
% Federated fine-tuning~\cite{zhang2024towards} emerges as a promising solution, enabling collaborative adaptation without exposing sensitive data.
% % Federated fine-tuning~\cite{zhang2024towards} offers a promising alternative, enabling collaborative LLM adaptation on edge devices without centralizing sensitive data. 
% Despite its promise, the high resource demands of LLMs severely bottleneck the deployment of federated fine-tuning on mobile devices~\cite{kuang2024federatedscope, wu2025learning}. 

Large Language Models (LLMs)~\cite{liu2024deepseek,bai2023qwen} are revolutionizing mobile intelligence, yet adapting them for downstream tasks remains a critical challenge. Centralized training is often rendered infeasible by privacy regulations that isolate task-specific data on user devices~\cite{mcmahan2017communication,wu2025breaking,wu2024heterogeneity,wu2024bridging}. While federated fine-tuning~\cite{zhang2024towards} offers a promising privacy-preserving solution for collaborative adaptation, its practical deployment on mobile devices is severely bottlenecked by the prohibitive resource demands of LLMs~\cite{tian2026floe,wudevelopmental,wu2025elastic,wu2025memory}.

% To mitigate the prohibitive resource overhead of full-parameter fine-tuning, numerous parameter-efficient federated fine-tuning methods have been developed~\cite{wu2025survey}, which freeze the base model and update only a lightweight, task-specific subset of parameters~\cite{bian2025survey}. Among them, adapter-based tuning~\cite{cai2022fedadapter} has garnered significant attention for its effectiveness. However, while these methods successfully reduce computational and communication overhead, they still fail to address the memory constraints of mobile devices. This is because they require loading the entire model into memory, a task made infeasible by the vast parameter size of LLMs. 
% For example, the LLaMA2-7B~\cite{llama2} can require up to 25GB of memory for its parameters, drastically exceeding the typical 4–12GB capacity of mobile devices~\cite{xu2023fwdllm}. Consequently, this memory wall prevents resource-constrained devices from participating, limiting the utilization of their valuable on-device data.

To mitigate the resource costs, parameter-efficient federated fine-tuning methods~\cite{wu2025survey}, particularly adapter-based approaches~\cite{cai2022fedadapter}, have been widely adopted. By freezing the backbone and updating only lightweight parameters~\cite{bian2025survey}, these methods successfully reduce computational and communication overheads. However, they fail to alleviate the fundamental memory constraint: the entire model must still be loaded into memory. For instance, LLaMA2-7B~\cite{llama2} requires approximately 25GB of memory, far exceeding the typical 4–12GB capacity of mobile devices~\cite{xu2023fwdllm}. This memory wall precludes resource-constrained devices from participating, leaving valuable on-device data unutilized.

To this end, we introduce \our, an innovative federated fine-tuning paradigm that breaks the memory wall via chain optimization. Instead of updating the LLM end-to-end, \our decouples the optimization into a series of sequential stages, with each stage dedicated to a single adapter. 
It commences by training the first adapter to convergence, freezes its parameters, and then triggers the training of the next one. This train-and-freeze process continues until all adapters are fully tuned. By focusing resources on one adapter at a time, \our drastically reduces peak memory usage, making federated fine-tuning feasible on resource-constrained devices. However, this novel paradigm poses several unique challenges, which we address through our core techniques.

$\bullet$ \textbf{Challenge 1: Representational Mismatch and Information Flow Bottleneck.}
The sequential nature of chain optimization inherently risks representational mismatch. As each adapter is optimized in isolation, it focuses solely on local objectives, neglecting the context of adjacent layers and creating semantic gaps. Furthermore, the train-and-freeze cycle impedes information flow: forward feature propagation is delayed until predecessor convergence, while backward gradient flow is confined to the active layer, preventing cross-layer co-adaptation. To address these, we propose Dynamic Layer Co-Tuning, which coordinates adjacent adapters to align feature spaces and facilitate inter-layer collaboration.

% The primary challenge of this chain optimization paradigm stems from its core design of training adapters sequentially and in isolation. This isolated process can lead to representational mismatch, as each adapter optimizes for its local layer without regard for adjacent layers, creating semantic gaps. Furthermore, the train-and-freeze cycle creates a bi-directional information flow bottleneck. In the forward pass, information propagation is delayed since subsequent layers only begin training after their predecessors converge. More critically, in the backward pass, gradients are confined to the single active adapter, which restricts the co-adaptation of parameters across the network. To resolve this, our Dynamic Layer Co-Tuning mechanism actively coordinates the fine-tuning of adjacent adapters, aligning their feature spaces to bridge semantic gaps and facilitating inter-layer collaboration.

$\bullet$ \textbf{Challenge 2: Short-Sighted Optimization Perspective.}
Sequential training also suffers from optimization myopia, as adapters are updated without error feedback from downstream layers. This exclusively local focus incentivizes over-specialization, causing the premature discard of generalizable information. Consequently, subsequent layers receive an impoverished feature space, degrading overall performance. To counteract this, we propose Globally Perceptive Optimization, which incorporates the model's holistic objective into each adapter's local loss. This mechanism forces adapters to balance local optimization with global utility, preserving critical information for synergistic, hierarchical learning.

% The sequential nature of this paradigm also gives rise to an optimization myopia. Since each adapter is updated without error feedback from downstream adapters, its optimization objective is purely local. This local objective incentivizes the adapter to learn highly specialized features, at the expense of discarding valuable, general-purpose information. This premature information loss creates an impoverished feature space for subsequent layers, limiting their learning capacity and undermining the model's final performance. To counteract this effect, we propose Globally Perceptive Optimization, which integrates the model's holistic learning objective directly into each adapter's local loss function. This forces each adapter to balance its immediate needs with the collective goal of the entire model, promoting the preservation of information and leading to more hierarchical and synergistic learning.

$\bullet$ \textbf{Challenge 3: Optimal Fine-Tuning Boundary Identification.} 
The hierarchical structure of LLMs, transitioning from low-level syntax to high-level semantics, raises a critical question: \textit{At which layer should the fine-tuning chain commence?} Initiating the chain too early is computationally wasteful, while starting too late risks insufficient adaptation. We resolve this dilemma with Function-Oriented Adaptive Tuning, which automatically identifies the optimal boundary by quantifying each layer's task contribution. By pinpointing the precise entry point, our method balances efficiency and effectiveness, retaining foundational knowledge while selectively tuning essential layers.

% Pre-trained LLMs exhibit a hierarchical functional structure, where lower layers capture basic linguistic patterns, while higher layers are specialized in semantic comprehension and task-specific reasoning.
% This functional hierarchy presents a critical question: At which layer should the chain of fine-tuning commence?
% Starting the chain too early is wasteful, as it unnecessarily retrains foundational knowledge. Conversely, starting too late may fail to adapt crucial intermediate features. To resolve this dilemma, we propose Function-Oriented Adaptive Tuning, which automatically identifies the optimal boundary by quantifying each layer's contribution to the task. By pinpointing the exact layer where adaptation should commence, our method ensures that we retain valuable pre-trained knowledge while efficiently tuning only the necessary layers.

Our main contributions are summarized as follows: 1) We introduce \our, an innovative federated fine-tuning paradigm that breaks the memory wall for LLM adaptation via chain optimization. 2) We identify and solve three core challenges in this paradigm with a suite of synergistic techniques. 3) Our extensive experiments show that \our significantly outperforms existing methods by a large margin.

\section{Related Work}

\subsection{Adapter-Based Federated Fine-Tuning}

Adapters~\cite{he2021effectiveness} have become a cornerstone of federated fine-tuning, with a significant body of work leveraging them to tackle key challenges such as data heterogeneity and slow convergence~\cite{zhangenhancing, yao2024federated}. For instance, to address data heterogeneity, C2A~\cite{kim2023client} generates personalized adapters via hypernetworks, while Fed-MNMT~\cite{liu2023communication} clusters devices to prevent negative transfer. Other approaches, such as FedAdapter~\cite{cai2023fedadapter}, focus on accelerating convergence through dynamic configurations. Despite these advancements, a fundamental limitation persists: the memory bottleneck. These methods require loading the entire model into memory, resulting in a prohibitive memory footprint. For instance, fine-tuning even the compact LLaMA3.1-3B~\cite{dubey2024llama} demands approximately 14GB of memory, exceeding the capacity of most mobile devices and rendering larger models (e.g., LLaMA2-7B) inaccessible. To address this memory wall, we propose \our, a chain optimization paradigm that minimizes peak memory usage, thereby enabling participation from resource-constrained edge devices.

\subsection{Memory-Aware Federated Fine-Tuning}

% Several approaches aim to mitigate memory costs during local training, primarily by reducing intermediate activations (e.g., FwdLLM~\cite{xu2023fwdllm}, FedKSeed~\cite{qin2023federated} using zeroth-order optimization) or minimizing trainable parameters (e.g., FLoRA~\cite{wang2024flora} via rank reduction). However, these methods overlook the dominant memory consumer: model parameters. For instance, in LoRA-based fine-tuning of LLaMA2-7B, base parameters account for 92.8\% of the total footprint, dwarfing the contributions of activations (7.2\%) and LoRA modules (0.018\%). Consequently, targeting these secondary components yields only marginal savings. In contrast, \our employs a chain optimization paradigm to directly address the primary bottleneck—the model parameters themselves.

Several approaches aim to mitigate memory costs during local training, primarily by reducing intermediate activations (e.g., FwdLLM~\cite{xu2023fwdllm} and FedKSeed~\cite{qin2023federated} using zeroth-order optimization) or minimizing trainable parameters (e.g., FLoRA~\cite{wang2024flora} via rank reduction). However, these methods overlook the dominant memory consumer: model parameters. For instance, in LoRA-based fine-tuning of LLaMA2-7B, base parameters account for 92.8\% of the total memory footprint, dwarfing the contributions of activations (7.2\%) and LoRA modules (0.018\%). Consequently, targeting these secondary components yields only marginal savings. In contrast, \our employs a chain optimization paradigm that directly addresses the primary bottleneck—the model parameters themselves.

% Several federated fine-tuning methods have been proposed to reduce the memory footprint during local training. These approaches attempt to minimize memory usage via two main strategies: reducing intermediate activations or shrinking the number of trainable parameters. 
% Methods like FwdLLM~\cite{xu2023fwdllm} and FedKSeed~\cite{qin2023federated} employ zeroth-order optimization to bypass backpropagation, thus eliminating activation storage. Others, like FLoRA~\cite{wang2024flora}, reduce the rank of LoRA modules to decrease the trainable parameter count.
% However, these methods suffer from a critical oversight: they ignore the primary source of memory consumption in LLMs, which is the storage of the model parameters themselves. For instance, during the LoRA-based fine-tuning of LLaMA2-7B~\cite{llama2}, the base model parameters account for approximately 92.8\% of the total memory footprint, while activations and the LoRA module constitute a mere 7.2\% and 0.018\%, respectively. Consequently, such approaches offer only marginal memory savings. Different from them, \our fine-tunes the model in a chain fashion, directly addressing the dominant memory footprint of model parameters.

\section{Preliminary and Motivation} 

\subsection{Basics of Adapter}

\begin{figure}[t]
  \centering
  \includegraphics[width=1\linewidth]{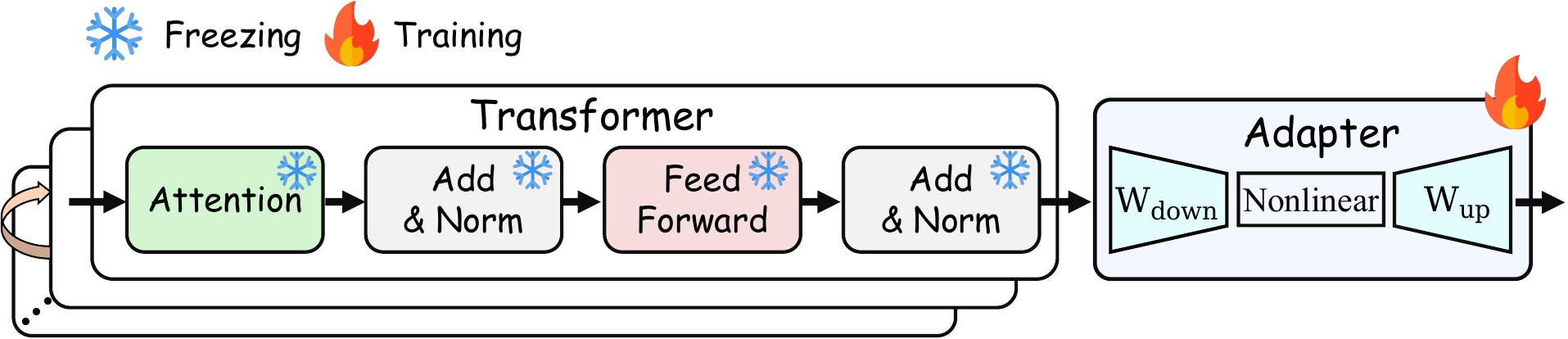}
  \caption{The configuration of the adapter.}
  \label{Adapter}
  \vspace{-3mm}
\end{figure}

The core principle of adapters is to keep pre-trained model parameters frozen while inserting lightweight, trainable modules into different parts of the model~\cite{houlsby2019parameter}. 
As illustrated in Figure~\ref{Adapter}, an adapter module operates via a bottleneck structure. It first projects the input hidden state $h$ into a low-dimensional space with a down-projection matrix (\( W_{down} \in \mathbb{R}^{u \times v} \)), applies a non-linear activation function \( f(\cdot) \), and then restores it to the original dimension using an up-projection matrix (\( W_{up} \in \mathbb{R}^{v \times u} \)). The output is then added to the input via a residual connection, formulated as:
\begin{equation}  
    \label{eq_adapter_trans}
    h \leftarrow h + f(h W_{down}) W_{up}.
\end{equation}
% For existing federated fine-tuning methods, participating devices update the full set of adapters simultaneously in each round. These adapter updates are then sent to the central server for aggregation, and the updated adapters are distributed back to devices for the next round. This iterative process is repeated until the global model converges.
% While the adapter significantly reduces computational and communication overhead, it still fails to address memory constraints. This is because during on-device training, the need to store the entire base model's weights, optimizer states, and intermediate activations in memory leads to a prohibitive memory footprint. This results in out-of-memory errors on resource-constrained devices, ultimately excluding them from the collaborative training process.

\subsection{Memory Wall in Federated Fine-Tuning}

This section motivates our work by quantifying the performance disparity between idealized and memory-constrained federated fine-tuning. We then analyze the memory footprint of adapter-based methods on representative LLMs, revealing critical insights that underpin the design of \our.

% \textbf{Observation 1: Memory constraints significantly degrade model performance.} Figure~\ref{fig:memory_wall} compares the performance of BERT under an idealized scenario (all devices participating) versus a practical one (where only devices with sufficient memory can participate). 
% We observe severe performance degradation for the practical setting across both IID and non-IID data distributions.
% For instance, on YELP-P~\cite{zhang2015character}, accuracy drops by 8.5\% (IID) and 11.8\% (non-IID). This decline is a direct result of the memory wall, which excludes low-end devices from federated training. These results underscore that memory constraints are not just a resource issue but a fundamental bottleneck to achieving high model performance.

\textbf{Observation 1: Memory constraints significantly degrade model performance.} Figure~\ref{fig:memory_wall} contrasts BERT's performance under idealized (full participation) versus practical (memory-constrained) scenarios. We observe severe performance degradation in the practical setting across both IID and non-IID distributions. On YELP-P~\cite{zhang2015character}, for instance, accuracy drops by 8.5\% (IID) and 11.8\% (non-IID). This decline stems directly from the memory wall, which systematically excludes low-end devices from training. These results underscore that memory constraints are not merely a resource hurdle but a fundamental bottleneck to model performance.

\definecolor{red}{RGB}{172,21,28}
\definecolor{blue}{RGB}{39,89,167}
\definecolor{red1}{RGB}{203,104,104}
\definecolor{blue1}{RGB}{104,155,203}
\definecolor{color1}{RGB}{235,164,122}
\definecolor{color3}{RGB}{78,172,183}
\definecolor{color2}{HTML}{95a792}

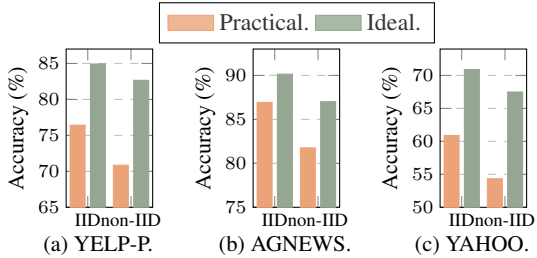
\begin{figure}[t]
\centering
\hspace{-2mm}
\begin{tikzpicture}
    \scriptsize{
  \begin{axis}[
    at={(-12em,-15.5em)},
    anchor=south west,
    ymajorgrids,
    grid style=dashed,
    legend style={at={(0.1,1)}, anchor=south west},
    legend cell align={left},
    ybar,
    enlarge x limits=0.5,
    xtick align=inside,
    height=.23\textwidth,
    width=.17\textwidth,
    bar width=0.8em,
    xlabel={\scriptsize{(a) YELP-P.}},
    xlabel style={scale=1.2, yshift=0.2em, xshift=-0.5em},
    ylabel=\footnotesize{\scriptsize Accuracy (\%)},
    ylabel style={scale=1.2, yshift=2.5em},
    symbolic x coords={{1}, {2},},
    xtick=data,
    ymin=65,
    ymax=87,
    ytick={65,70,75,80,85},
    nodes near coords align={vertical},
    xticklabels={IID, non-IID},
    ylabel style={yshift=-2.7em},
    yticklabel style={/pgf/number format/fixed,/pgf/number format/fixed zerofill,/pgf/number format/precision=0,rotate=0,scale=1.0},
    legend style={yshift=0.3em,xshift=4.5em,font={\tiny},cells={anchor=west},fill opacity=0.8, scale=0.5, legend columns=3, font=\small}
    ]
    \addplot[fill=color1, draw=color1, area legend] coordinates {({1},76.42) ({2},70.85)};
    \addlegendentry{\scalebox{1.0}{{Practical.}}}
    \addplot[fill=color2, draw=color2, area legend] coordinates {({1},84.92) ({2},82.65)};
    \addlegendentry{\scalebox{1.0}{Ideal.}}
    % \addplot[fill=color3, draw=color3, area legend] coordinates {({1},89.5) ({2},88.4)};
    % \addlegendentry{\scalebox{1.0}{\textsc{TheoFL}}}
  \end{axis}
	
  \begin{axis}[
    at={(-2em,-15.5em)},
    anchor=south west,
    ymajorgrids,
    grid style=dashed,
    legend style={at={(0.02,1)}, anchor=south west},
    legend cell align={left},
    ybar,
    enlarge x limits=0.5,
    xtick align=inside,
    height=.23\textwidth,
    width=.17\textwidth,
    bar width=0.8em,
    xlabel={\scriptsize{(b) AGNEWS.}},
    xlabel style={scale=1.2, yshift=0.2em, xshift=-0.5em},
    ylabel=\footnotesize{\scriptsize Accuracy (\%)},
    ylabel style={scale=1.2, yshift=2.5em},
    symbolic x coords={{1}, {2},},
    xtick=data,
    ymin=75,
    ymax=93,
    ytick={75,80,85,90},
    nodes near coords align={vertical},
    xticklabels={IID, non-IID},
    ylabel style={yshift=-2.7em},
    yticklabel style={/pgf/number format/fixed,/pgf/number format/fixed zerofill,/pgf/number format/precision=0,rotate=0,scale=1.0},
    legend style={yshift=0.2em,xshift=4.2em,font={\tiny},cells={anchor=west},fill opacity=0.8, scale=1.0, legend columns=3}
    ]
    \addplot[fill=color1, draw=color1, area legend] coordinates {({1},86.93) ({2},81.76)};
    % \addlegendentry{\scalebox{1.0}{\textsc{FedAvg}}}
    \addplot[fill=color2, draw=color2, area legend] coordinates {({1},90.14) ({2},87.01)};
    % \addlegendentry{\scalebox{1.0}{\textsc{Sequential}}}
    % \addplot[fill=color3, draw=color3, area legend] coordinates {({1},59.6) ({2},58.1)};
    % \addlegendentry{\scalebox{1.0}{\textsc{TheoFL}.}}
  \end{axis}

  \begin{axis}[
    at={(8em,-15.5em)},
    anchor=south west,
    ymajorgrids,
    grid style=dashed,
    legend style={at={(0.02,1)}, anchor=south west},
    legend cell align={left},
    ybar,
    enlarge x limits=0.5,
    xtick align=inside,
    height=.23\textwidth,
    width=.17\textwidth,
    bar width=0.8em,
    xlabel={\scriptsize{(c) YAHOO.}},
    xlabel style={scale=1.2, yshift=0.2em, xshift=-0.5em},
    ylabel=\footnotesize{\scriptsize Accuracy (\%)},
    ylabel style={scale=1.2, yshift=2.5em},
    symbolic x coords={{1}, {2},},
    xtick=data,
    ymin=50,
    ymax=74,
    ytick={50,55,60,65,70},
    nodes near coords align={vertical},
    xticklabels={IID, non-IID},
    ylabel style={yshift=-2.7em},
    yticklabel style={/pgf/number format/fixed,/pgf/number format/fixed zerofill,/pgf/number format/precision=0,rotate=0,scale=1.0},
    legend style={yshift=0.2em,xshift=4.2em,font={\tiny},cells={anchor=west},fill opacity=0.8, scale=1.0, legend columns=3}
    ]
    \addplot[fill=color1, draw=color1, area legend] coordinates {({1},60.89) ({2},54.33)};
    % \addlegendentry{\scalebox{1.0}{\textsc{FedAvg}}}
    \addplot[fill=color2, draw=color2, area legend] coordinates {({1},70.89) ({2},67.48)};
    % \addlegendentry{\scalebox{1.0}{\textsc{Sequential}}}
    % \addplot[fill=color3, draw=color3, area legend] coordinates {({1},59.6) ({2},58.1)};
    % \addlegendentry{\scalebox{1.0}{\textsc{TheoFL}.}}
  \end{axis}
  
}   
\end{tikzpicture}
\vspace{-3mm}
% \caption{Performance comparison between practical deployment and ideal conditions on BERT.}
\caption{Performance comparison of BERT fine-tuning under practical versus idealized deployment conditions.}
\label{fig:memory_wall}
\vspace{-3mm}
\end{figure}

\begin{figure}[t]
  \centering
  \includegraphics[width=1\linewidth]{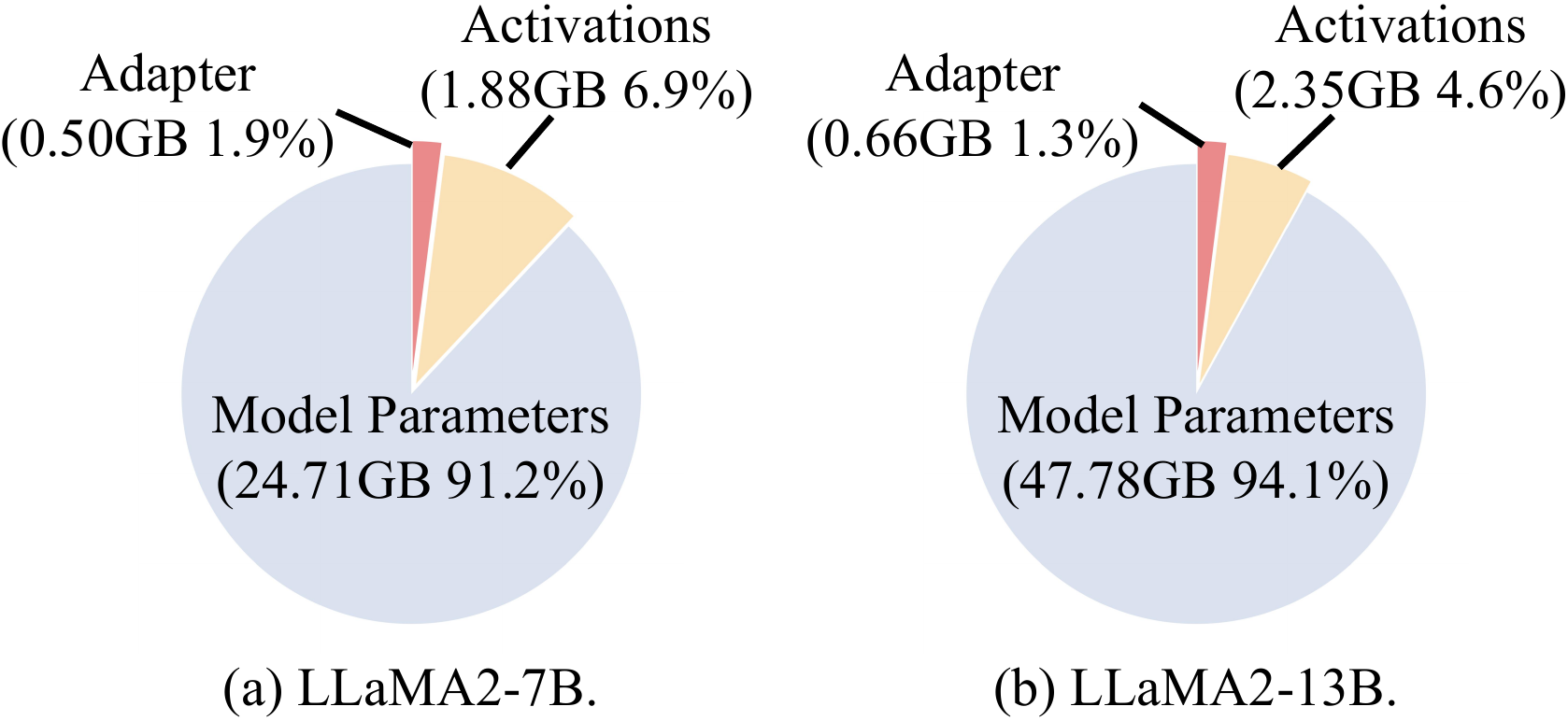}
  \caption{Memory usage breakdown during adapter-based fine-tuning. The ``Adapter'' component includes its parameters and gradients.}
  \label{fig_memory_breakdown}
  \vspace{-4mm}
\end{figure}

\textbf{Observation 2: Adapter-based tuning fails to scale with modern LLMs.} 
We next investigate whether adapters are sufficient to overcome the memory wall. To verify this, we profile the memory usage of fine-tuning contemporary LLMs with adapters. The results in Figure~\ref{fig_memory_breakdown} show that even with adapters, fine-tuning LLaMA2-7B still requires approximately 27.1GB of memory. This demand escalates sharply with model scale, reaching 50.8GB for LLaMA2-13B, a requirement far beyond the capacity of any consumer edge device. This confirms that standard adapter-based tuning, on its own, is an inadequate solution for bringing large-scale federated fine-tuning to the edge.

\textbf{Observation 3: Model parameters overwhelmingly dominate the memory footprint.} Our memory breakdown analysis identifies the base model parameters as the primary bottleneck. As shown in Figure~\ref{fig_memory_breakdown}, for LLaMA2-7B, parameters consume 91.2\% of the total memory, dwarfing intermediate activations (6.9\%) and adapter modules (1.9\%). This dominance intensifies with scale, reaching 94.1\% for LLaMA2-13B. This finding exposes a critical limitation in existing techniques that target activations or trainable parameters: since these components constitute a negligible fraction of the total footprint, their potential savings are inherently marginal. To truly overcome the memory bottleneck, an effective strategy must fundamentally reduce the number of model parameters residing in memory during fine-tuning.

\section{Our Method: \our}

% Motivated by the preceding analysis, we introduce \our, a novel federated fine-tuning framework designed to break the memory wall via chain optimization. It further integrates three synergistic techniques to ensure effectiveness and robustness: 1) Dynamic Layer Co-Tuning to maintain representational coherence and information flow; 2) Globally Perceptive Optimization to counteract local optimization myopia; and 3) Function-Oriented Adaptive Tuning to automatically identify the optimal starting point of the chain.

\subsection{The Chain Optimization Paradigm}

\begin{figure}[!t]
  \centering
  \includegraphics[width=1\linewidth]{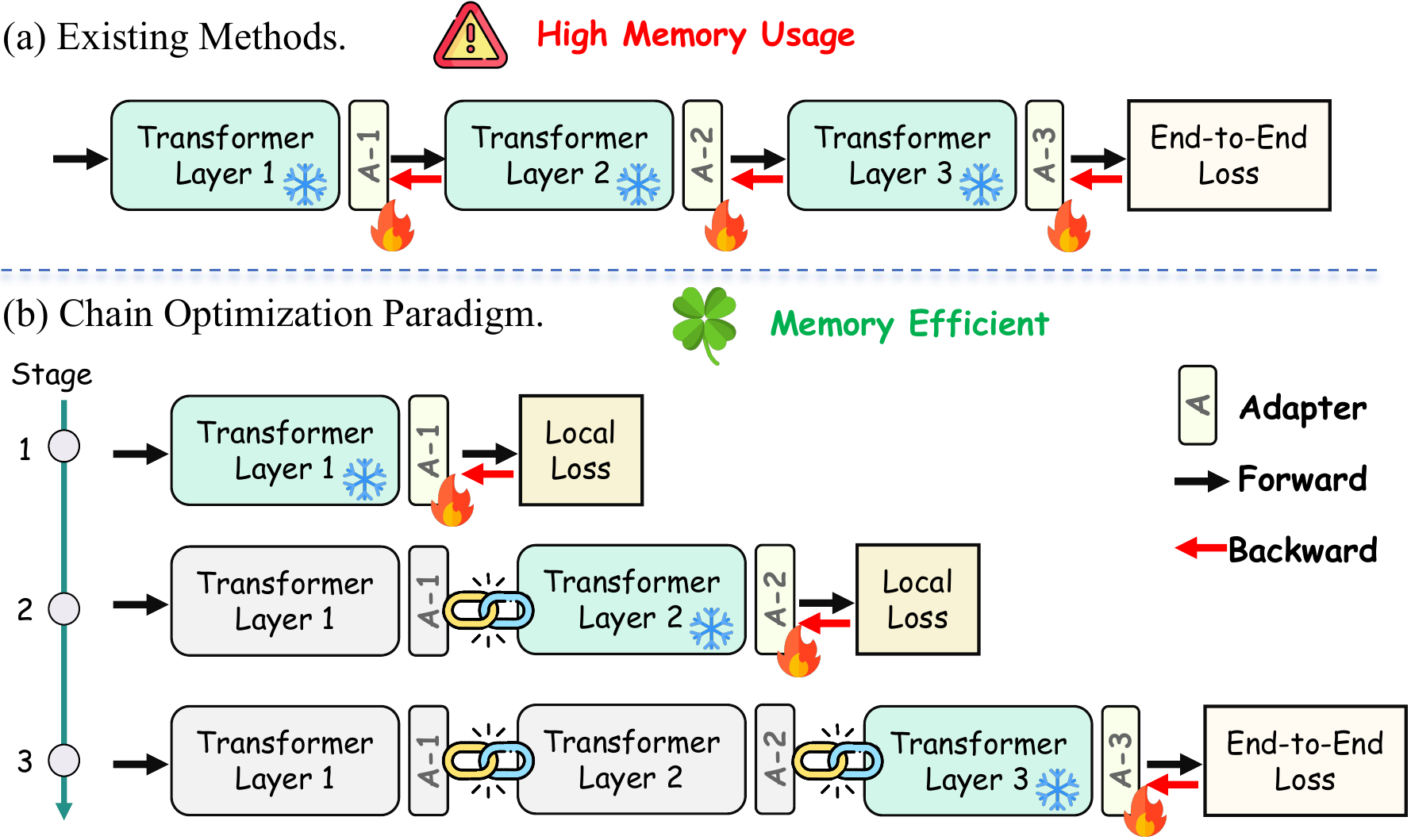}
  \caption{Comparison of existing methods with our chain optimization paradigm on a three-layer transformer model. The local loss is computed by attaching an output layer to each adapter, while the end-to-end loss is derived from the final output of the entire model.}
  \label{framework}
  \vspace{-4mm}
\end{figure}

Figure~\ref{framework}(b) illustrates the chain optimization paradigm. Unlike conventional methods that update all adapters end-to-end (Figure~\ref{framework}(a)), our approach decomposes the fine-tuning process into sequential stages, each dedicated to a specific layer and its adapter. This process begins by fine-tuning the first adapter, which is subsequently frozen upon convergence. The model then incorporates the second layer and its adapter, initiating the second training stage. This train-and-freeze process continues until all adapters are fully optimized, gradually expanding the model to its complete architecture. 

% Figure~\ref{framework}(b) illustrates the chain optimization paradigm. In contrast to existing methods that update all adapters end-to-end (Figure~\ref{framework}(a)), our paradigm deconstructs the fine-tuning process into a sequence of distinct stages. Each stage is dedicated to a single layer and its corresponding adapter. 
% The process begins by fine-tuning the first adapter, which is subsequently frozen upon convergence. The model then incorporates the second layer and its adapter, initiating the second training stage. This train-and-freeze process continues until all adapters are fully optimized, gradually expanding the model to its complete architecture. 

In each stage, only the corresponding adapter undergoes parameter updates, eliminating the storage overhead for gradients and optimizer states of inactive layers. 
Specifically, preceding layers operate in inference mode, allowing immediate release of memory after the forward pass, while subsequent layers remain idle to prevent unnecessary allocation. However, this paradigm introduces three unique challenges, which we address below.

% Specifically, preceding layers operate in inference mode, allowing immediate release of intermediate activations, while subsequent layers remain idle to prevent unnecessary allocation. However, this paradigm introduces three unique challenges, which we address below.

\subsection{Dynamic Layer Co-Tuning}\label{sec_DLCT}

The sequential design of chain optimization inherently induces representational mismatch and bottlenecks bi-directional information flow. First, isolated training drives adapters to over-specialize for their local layers, neglecting adjacent context and causing semantic gaps. Second, the protocol disrupts information exchange: forward propagation is stalled until predecessor convergence, while backward gradients are confined to the active layer, precluding cross-layer co-adaptation.

% The sequential, one-at-a-time design of the chain optimization paradigm directly leads to representational mismatch between adjacent layers and a bi-directional information flow bottleneck.
% First, when an adapter is trained alone, it becomes highly specialized for its immediate layer, with no awareness of the needs of the subsequent layer. This localized focus can cause the feature representations of adjacent layers to drift apart, creating semantic gaps.
% Second, in the forward pass, timely information propagation is hindered because downstream layers can only begin training after their predecessors have fully converged. More critically, in the backward pass, gradients are trapped within the single active adapter, preventing layers from co-adapting and learning from each other.

\begin{figure}[t]
  \centering
  \includegraphics[width=1\linewidth]{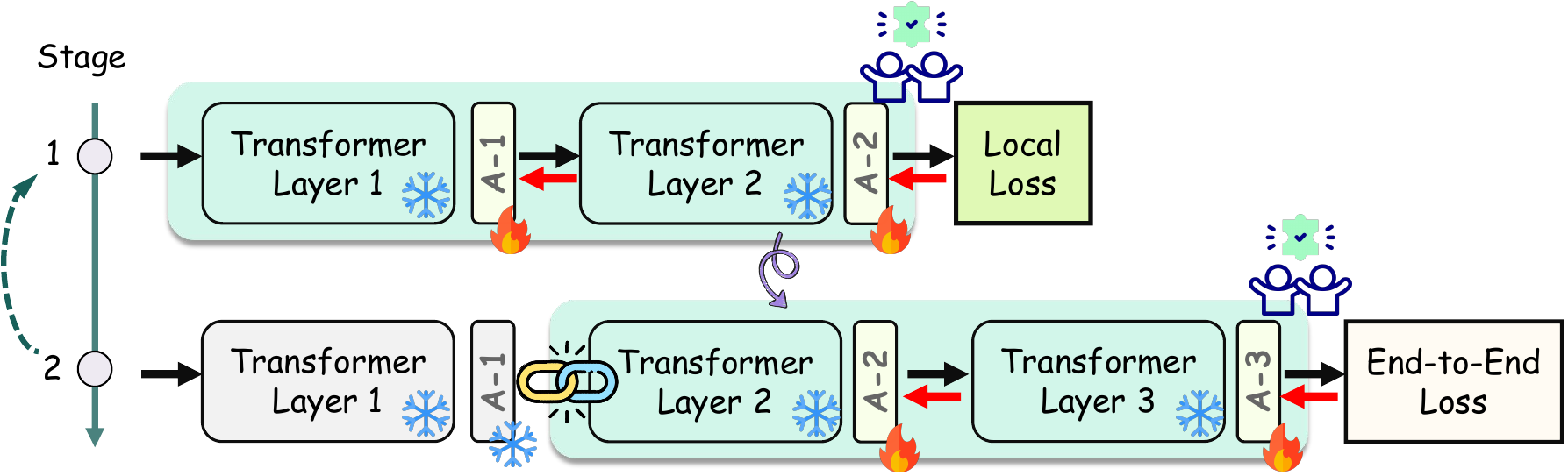}
  \caption{Overview of the Dynamic Layer Co-Tuning mechanism, illustrated with an example where the sliding window size is two ($Q=2$).}
  \label{component1}
  \vspace{-3mm}
\end{figure}

To address these issues, we propose the Dynamic Layer Co-Tuning (DLCT) to foster inter-layer collaboration. Instead of fine-tuning adapters in isolation, DLCT creates a sliding window that moves sequentially along the chain. Adapters within this window are co-tuned simultaneously. The size of the sliding window is $Q$. When advancing to the next stage, the window then slides forward by one layer, creating an overlap of $Q-1$ adapters between consecutive stages. 
For instance, with $Q=2$ (Figure~\ref{component1}), the first stage co-tunes adapters 1 and 2; subsequently, the window shifts to adapters 2 and 3. This overlap allows adapter 2 to be further refined alongside adapter 3, effectively bridging the optimization context between stages.

% For instance, as shown in Figure~\ref{component1}, when $Q=2$, two adapters are co-tuned in each stage. In the first stage, adapters 1 and 2 are fine-tuned together; in the second stage, adapter 2, having already benefited from insights gained during first stage, is further refined alongside adapter 3.

This co-tuning mechanism offers two key advantages:
1) \textbf{Bridging Semantic Gaps.} The inter-stage overlap directly mitigates representational mismatch. By training an adapter alongside both its upstream and downstream neighbors, it is incentivized to function as a semantic anchor, aligning feature representations across layers to ensure coherence.
2) \textbf{Breaking Gradient Isolation.} DLCT enables gradient information to freely propagate across adapters using its sliding window. As shown in Figure~\ref{component1}, the shared adapter (e.g., Adapter 2) acts as a conduit: it receives gradients from Adapter 3 in the second stage and influences Adapter 1 from the first, effectively linking non-adjacent layers. Furthermore, to enhance information flow, the window advances continuously in each round rather than waiting for stage-wise convergence. The process cycles iteratively, allowing for multiple passes of holistic refinement across the model.

\subsection{Globally Perceptive Optimization}

While DLCT ensures representational coherence and facilitates cross-layer information flow, the optimization perspective remains inherently myopic. Lacking feedback from downstream layers, each adapter greedily maximizes its immediate local objective. This short-sighted behavior induces the premature loss of valuable  information, resulting in an impoverished feature space for subsequent layers and ultimately degrading overall performance.

% While DLCT ensures representational coherence and facilitates cross-layer information flow, the optimization perspective at each stage remains "short-sighted," as the absence of feedback from downstream layers causes each adapter to greedily learn features that maximize its immediate, local objective. This myopic behavior leads to the premature loss of valuable, general-purpose information, creating an impoverished feature space for subsequent layers and ultimately degrading the model's overall performance~\cite{wu2024neulite}.

To endow adapters with optimization foresight, we introduce Globally Perceptive Optimization (GPO). This strategy integrates the model's holistic objective into local updates, compelling adapters to align with global goals. To achieve this, we design an auxiliary output branch to compute the global loss. A naive approach would pass the current hidden state through all subsequent layers; however, this would incur substantial computation overhead and memory consumption. To address this, we propose a lightweight alternative: our auxiliary branch comprises only the subsequent adapters and the final output layer. This design leverages adapters as compact, low-rank approximations of layer transformations, enabling accurate estimation of the end-to-end loss with minimal overhead.

\begin{figure}[t]
  \centering
  \includegraphics[width=1\linewidth]{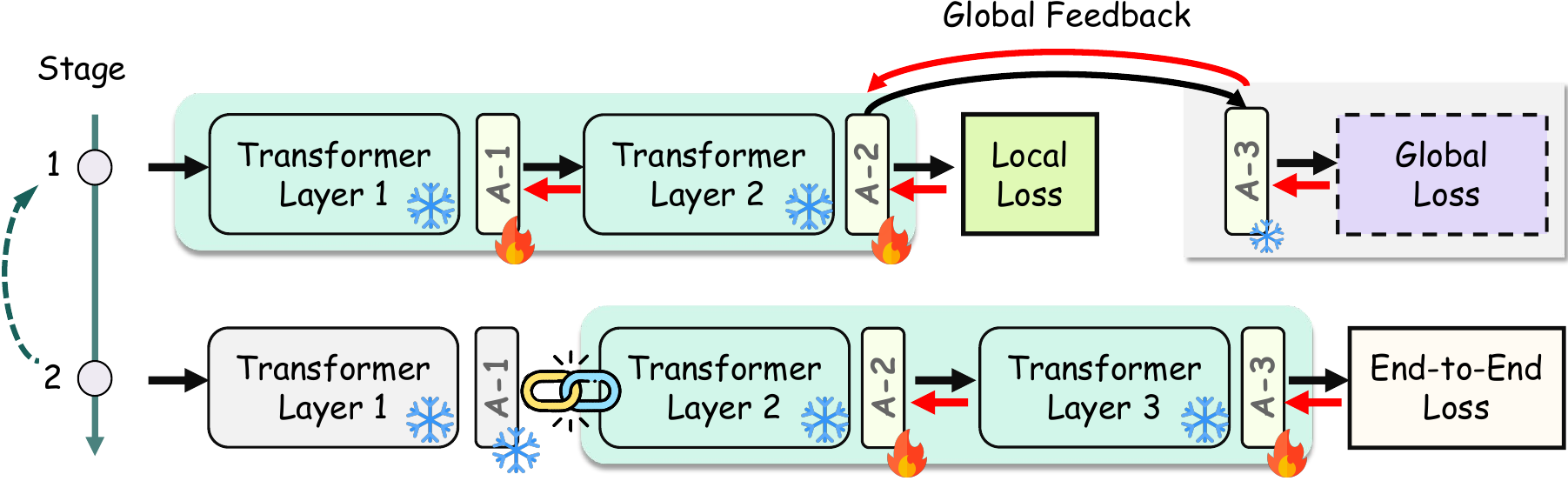}
  \caption{The Globally Perceptive Optimization strategy. An auxiliary output branch, composed only of subsequent adapters and the final output layer, is used to compute a global loss signal.}
  \label{component2}
  \vspace{-3mm}
\end{figure}

As illustrated in Figure~\ref{component2}, the optimization at each stage is guided by a dual-loss signal. For example, when co-tuning adapters 1 and 2, the training objective combines the local loss from adapter 2's output with the global loss computed via the lightweight auxiliary branch. Except for the final stage (which uses only the end-to-end loss), the objective at each stage $m$ is formulated as:
\begin{equation}
    \text{Loss}_{m} = \text{Local Loss} + \lambda \cdot \text{Global Loss},
    \label{update_objective}
\end{equation}
where $\lambda$ is a hyperparameter balancing local and global objectives. By incentivizing adapters to learn features that are locally optimal yet globally beneficial, GPO promotes a synergistic and foresight-driven optimization process.

\subsection{Function-Oriented Adaptive Tuning}

While DLCT and GPO optimize training dynamics, the inherent functional hierarchy of LLMs, transitioning from shallow syntax to deep semantics, poses a strategic question: \textit{At which layer should the fine-tuning chain commence?} Indiscriminate full-chain tuning is computationally redundant and risks degrading general-purpose representations. Conversely, selectively tuning task-critical layers enhances both efficiency and performance. However, pinpointing the optimal boundary between general and task-specific knowledge is non-trivial, particularly given the data heterogeneity.

% While DLCT and GPO optimize training dynamics, the inherent functional hierarchy of LLMs, progressing from shallow syntax to deep semantics, raises a critical question: \textit{At which layer should the fine-tuning chain commence?} Fine-tuning the entire chain is computationally redundant and risks degrading general-purpose representations. Conversely, selectively tuning task-critical layers can enhance efficiency and performance. However, pinpointing the optimal boundary between general and task-specific knowledge is non-trivial, particularly given the data heterogeneity in federated settings.

\begin{figure}[t]
  \centering
  \includegraphics[width=1\linewidth]{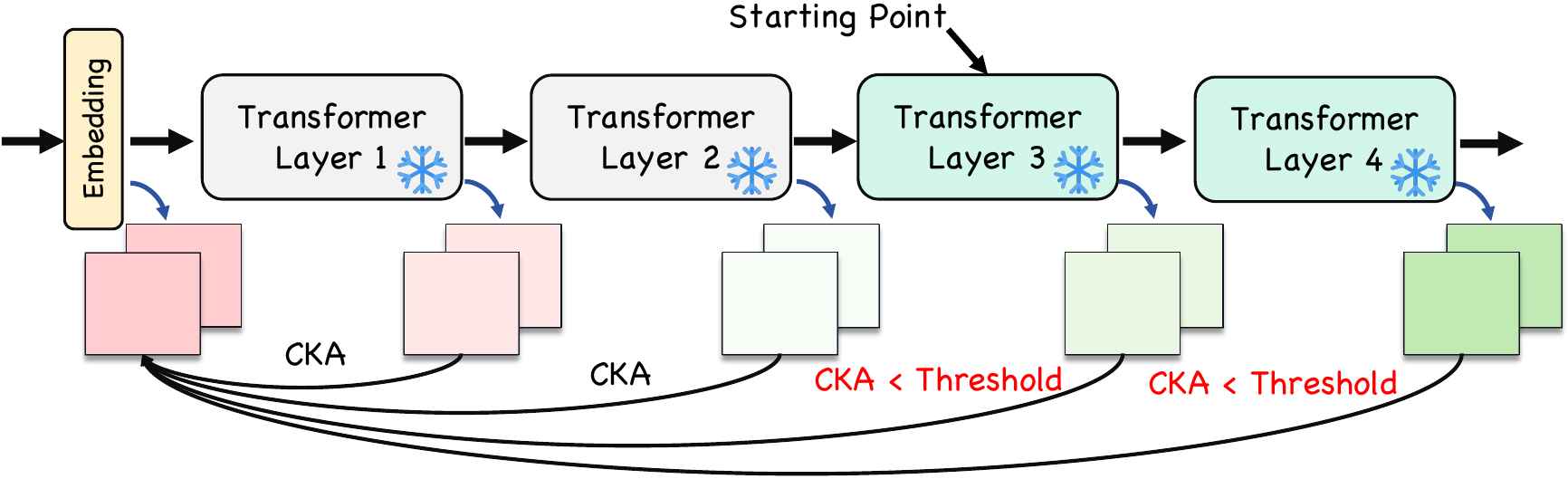}
  \caption{The Function-Oriented Adaptive Tuning scheme. Layer functionality is analyzed using CKA.}
  \label{component3}
  \vspace{-3mm}
\end{figure}

To address this, we propose Function-Oriented Adaptive Tuning (FOAT), a data-driven scheme that automatically pinpoints the optimal fine-tuning entry point. Our approach is grounded in the insight that the functional role of a layer correlates with its feature transformation intensity. Layers exhibiting minimal deviation from the initial input are considered general-purpose and kept frozen. Conversely, layers that induce significant feature divergence are deemed task-specific, marking them as prime candidates for adaptation.

% To address this challenge, we propose the Function-Oriented Adaptive Tuning (FOAT) scheme. FOAT provides a data-driven method to automatically pinpoint the optimal starting point for the fine-tuning chain. Our key insight is that this optimal boundary can be identified by measuring how much each layer transforms the initial input features. Layers that perform minimal transformation, meaning their output features are highly similar to the initial input, are general-purpose and should remain frozen. Conversely, layers that significantly alter the features are deemed task-specific and thus prime candidates for fine-tuning.

To quantify feature transformation, FOAT employs Centered Kernel Alignment (CKA)~\cite{kornblith2019similarity} to assess the similarity between layer-wise representations and the initial input. As illustrated in Figure~\ref{component3}, this process initializes before federated training. Each participating device performs a single forward pass with the global model on local data to extract activations, computing the CKA similarity scores as follows:
% To quantify this degree of feature transformation, FOAT leverages Centered Kernel Alignment (CKA)~\cite{kornblith2019similarity} to measure the similarity between each layer's output activations and the initial input. The process, illustrated in Figure~\ref{component3}, begins before the federated training starts. Each device receives the global model, performs a single forward pass on its local data to collect layer-wise activations, and computes the CKA similarity scores, which are formulated as:
\begin{equation}
    \label{eq_CKA}
    \text{CKA}(Z_{i}, Z_{j}) = \frac{\text{HSIC}(Z_{i}, Z_{j})}{\sqrt{\text{HSIC}(Z_{i}, Z_{i}) \cdot \text{HSIC}(Z_{j}, Z_{j})}},
\end{equation}
where $Z_{i}$ and $Z_{j}$ denote layer activations and HSIC is the Hilbert-Schmidt Independence Criterion~\cite{gretton2005measuring} (detailed in Appendix~\ref{appendix_hsic}).
Following local computation, devices upload CKA scores to the server for aggregation, revealing the global feature evolution from general to task-specific. The server then identifies the optimal starting layer, $L_{start}$, as the first layer where the aggregated CKA value falls below a predefined threshold $T$. This data-driven, inference-only strategy is inherently robust to non-IID distributions and ensures efficiency by preserving foundational knowledge while targeting task-critical layers. The complete workflow and convergence analysis are provided in Appendices~\ref{appendix_algo} and~\ref{appendix_convergence}, respectively.

\section{Experiments}

\subsection{Experimental Setup}

\noindent\textbf{Models and Datasets.} We evaluate \our on both text classification and instruction tuning tasks. For text classification, we employ three models of varying complexity: DistilBERT-base~\cite{sanh2019distilbert}, BERT-base~\cite{devlin2018bert}, and RoBERTa-large~\cite{liu2019roberta}. These are benchmarked on four datasets: YELP-P (binary sentiment)~\cite{zhang2015character}, AGNEWS (4-class news)~\cite{zhang2015character}, YAHOO (10-class topic)~\cite{zhang2015character}, and 20NEWS (20-class news)~\cite{lang1995newsweeder}. For instruction tuning, we fine-tune LLaMA2-7B~\cite{llama2} and LLaMA3.1-8B~\cite{dubey2024llama} using the Alpaca-GPT4 dataset~\cite{peng2023instruction}. Evaluation covers a diverse suite of benchmarks: MMLU~\cite{hendrycks2020measuring} for knowledge understanding, BBH~\cite{bbh} and DROP~\cite{Dua2019DROP} for complex reasoning, and CRASS~\cite{frohberg2022crass} for counterfactual reasoning~\cite{ye2024openfedllm}.

% \noindent\textbf{Models and Datasets.} To thoroughly evaluate the effectiveness of \our, we conduct experiments on both text classification and instruction tuning tasks. For text classification, we use three widely used transformer-based models with varying complexities: 1) DistilBERT-base~\cite{sanh2019distilbert}, 2) BERT-base~\cite{devlin2018bert}, and 3) RoBERTa-large~\cite{liu2019roberta}. 
% The benchmark datasets include: 1) YELP-P~\cite{zhang2015character}, a binary sentiment dataset, 2) AGNEWS~\cite{zhang2015character}, a 4-class news topic dataset, 3) YAHOO~\cite{zhang2015character}, a 10-class topic dataset, and 4) 20NEWS~\cite{lang1995newsweeder}, a 20-class news topic dataset.
% For instruction tuning, we fine-tune LLaMA2-7B~\cite{llama2} and LLaMA3.1-8B~\cite{dubey2024llama} on the Alpaca-GPT4 dataset~\cite{peng2023instruction}.
% Evaluation is conducted across a diverse set of benchmarks: MMLU~\cite{hendrycks2020measuring} for knowledge understanding, BBH~\cite{bbh} and DROP~\cite{Dua2019DROP} for complex reasoning, and CRASS~\cite{frohberg2022crass} for counterfactual reasoning~\cite{ye2024openfedllm}.

\noindent\textbf{Data Distribution and Scale.} We evaluate model performance under both IID and non-IID settings, generating non-IID partitions via a Dirichlet distribution with $\alpha=1$~\cite{zhang2026coupled,zhang2025subspace}. To assess system scalability, we vary the total client population across datasets: 20 for Alpaca-GPT4, 100 for 20NEWS, 1,000 for AGNEWS/YELP-P, and 10,000 for YAHOO. Further implementation details and metrics are provided in Appendix~\ref{appendx_implementation_details}.

\definecolor{steelbluev2}{HTML}{DAE8FC}
\definecolor{steelblue}{HTML}{82B0D2}
\definecolor{my_c1}{HTML}{c6f1e7}
\definecolor{my_c2}{HTML}{13829b}

\begin{table*}[t]

\vspace{-2mm}
\renewcommand{\arraystretch}{1} 
\resizebox{1\textwidth}{!}{ % 保持表格自适应宽度

  \centering  
    \begin{tabular}{clccccccccc}
    \toprule[1pt]
    &\multirow{2}{*}{\textbf{Method}} &  \multicolumn{2}{c}{\textbf{YELP-P}} & \multicolumn{2}{c}{\textbf{AGNEWS}}  & \multicolumn{2}{c}{\textbf{YAHOO}} & \multicolumn{2}{c}{\textbf{20NEWS}} & \multirow{2}{*}{\textbf{Average}}  \\
    \cmidrule(lr){3-4}
    \cmidrule(lr){5-6}
    \cmidrule(lr){7-8}
    \cmidrule(lr){9-10}

    && \textbf{IID} & \textbf{non-IID}& \textbf{IID} & \textbf{non-IID}& \textbf{IID} & \textbf{non-IID}& \textbf{IID} & \textbf{non-IID}&  \\
    \midrule[1pt]

    &\multicolumn{10}{c}{\textbf{DistilBERT-base}} \\ 
    \midrule[1pt]
    \rowcolor{gray!10}
    \textbf{Lower Bound}& No-FT$^{\dagger}$ & 50.04 & 50.04 & 25.13 & 25.13 & 10.05 & 10.05 & 5.01 & 5.01 & / \\
    \midrule
    
    \multirow{3}{*}{\textbf{\shortstack{Memory\\Unaware}}}&Linear Probing~\cite{kornblith2019better} & 71.56 & 67.53 & 85.76 & 82.84 & 59.22 & 57.13 & 73.74 & 69.83  & 70.95 \textcolor{my_c2}{($\downarrow$ 11.50)}\\ 
    &FedAdapter~\cite{cai2022fedadapter}                  
                            & 82.16 & 78.57 & 89.00 & 84.41 & 68.07 & 64.82 & 76.29 & 72.89 &77.03 \textcolor{my_c2}{($\downarrow$ 5.42)} \\ 
    & C2A~\cite{kim2023client}                       
                            & 80.24 & 76.95 & 87.49 & 82.88 & 66.24 & 61.87 & 74.65 & 70.47 &75.10 \textcolor{my_c2}{($\downarrow$ 7.35)} \\
                            
    \midrule
    
    \multirow{5}{*}{\textbf{\shortstack{Memory\\Aware}}}&FwdLLM~\cite{xu2023fwdllm}                      
                            & 82.00 & 77.61 & 88.79 & 83.03 & 65.77 & 61.93 & 63.70 & 59.34 &72.77 \textcolor{my_c2}{($\downarrow$ 9.68)} \\ 
    &FedKSeed~\cite{qin2023federated}
    						& 81.91 & 77.42 & 88.51 & 82.86 & 66.92 & 63.11 & 71.49 & 67.83 &75.01 \textcolor{my_c2}{($\downarrow$ 7.44)} \\
    &FLoRA~\cite{wang2024flora}
    						& 82.07 & 77.52 & 88.65 & 82.93 & 67.00 & 63.12 & 72.00 & 67.91 &75.15 \textcolor{my_c2}{($\downarrow$ 7.30)} \\
    &FedRA~\cite{su2024fedra}                       
                            & 81.55 & 77.91 & 88.92 & 84.25 & 68.36 & 65.03 & 76.66 & 72.71 &76.92  \textcolor{my_c2}{($\downarrow$ 5.53)} \\
    &\our\cellcolor{my_c1!50}       & \textbf{86.57}\cellcolor{my_c1!50} & \textbf{84.22}\cellcolor{my_c1!50} & \textbf{92.57}\cellcolor{my_c1!50} & \textbf{90.18}\cellcolor{my_c1!50} & \textbf{73.78}\cellcolor{my_c1!50} & \textbf{70.39}\cellcolor{my_c1!50} & \textbf{82.29}\cellcolor{my_c1!50} & \textbf{79.63}\cellcolor{my_c1!50} & \textbf{82.45}\cellcolor{my_c1!50}\\

    \midrule
    \textbf{Upper Bound}&Full Adapters$^{\dagger}$ & \underline{84.76} & \underline{82.58} & \underline{90.05} & \underline{86.96} & \underline{70.85} & \underline{67.42} & \underline{78.66} & \underline{75.45} &\underline{79.59} \textcolor{my_c2}{($\downarrow$ 2.86)}\\

    \midrule[1pt]

    &\multicolumn{10}{c}{\textbf{BERT-base}} \\ % 3表示跨越3列，居中对齐
    \midrule[1pt]

    \rowcolor{gray!10}
    \textbf{Lower Bound}& No-FT$^{\dagger}$ & 49.87 & 49.87 & 24.99 & 24.99 & 9.94 & 9.94 & 5.05 & 5.05 & / \\

    \midrule

    \multirow{3}{*}{\textbf{\shortstack{Memory\\Unaware}}}&Linear Probing~\cite{kornblith2019better} & 69.09 & 64.96 & 79.88 & 71.55 & 55.23 & 52.82 & 67.62 & 60.97 & 65.27 \textcolor{my_c2}{($\downarrow$ 16.85)} \\ 
    
    &FedAdapter~\cite{cai2022fedadapter}                  
                            & 77.84 & 72.59 & 88.41 & 84.03 & 62.67 & 59.74 & 69.42 & 65.76 & 72.56 \textcolor{my_c2}{($\downarrow$ 9.56)} \\ 
    & C2A~\cite{kim2023client}                       
                            & 76.39 & 71.16 & 86.81 & 82.11 & 60.78 & 55.73 &68.14  & 61.92  & 70.38 \textcolor{my_c2}{($\downarrow$ 11.74)}\\
                            
    \midrule
    
    \multirow{5}{*}{\textbf{\shortstack{Memory\\Aware}}}&FwdLLM~\cite{xu2023fwdllm}                      
                            & 81.86 & 77.38 & 87.47 & 80.68 & 66.05 & 62.16 & 5.21$^*$ & 5.01$^*$ &  58.23 \textcolor{my_c2}{($\downarrow$ 23.89)} \\ 
    &FedKSeed~\cite{qin2023federated}
    						& 82.13 & 77.61 & 87.81 & 82.23 & 67.05 & 63.21 & 71.92 & 68.01 &75.00 \textcolor{my_c2}{($\downarrow$ 7.12)} \\
    &FLoRA~\cite{wang2024flora}
    						& 82.19 & 77.75 & 87.98 & 82.51 & 67.12 & 63.24 & 72.03 & 68.15 &75.12 \textcolor{my_c2}{($\downarrow$ 7.00)} \\
    &FedRA~\cite{su2024fedra}                       
                            & 82.26 & 78.67 & 88.04 & 83.07 & 68.97 & 65.58 & 77.45 & 73.47 & 77.19 \textcolor{my_c2}{($\downarrow$ 4.93)} \\
    &\our\cellcolor{my_c1!50}       & \textbf{86.40}\cellcolor{my_c1!50} & \textbf{84.05}\cellcolor{my_c1!50} & \textbf{91.89}\cellcolor{my_c1!50} & \textbf{89.67}\cellcolor{my_c1!50} & \textbf{73.65}\cellcolor{my_c1!50} & \textbf{70.17}\cellcolor{my_c1!50} & \textbf{82.21}\cellcolor{my_c1!50} & \textbf{78.94}\cellcolor{my_c1!50} & \textbf{82.12}\cellcolor{my_c1!50} \\
    
    \midrule
    \textbf{Upper Bound}&Full Adapters$^{\dagger}$ & \underline{84.92} & \underline{82.65} & \underline{90.14} & \underline{87.01} & \underline{70.89} & \underline{67.48} & \underline{78.69} & \underline{75.46} & \underline{79.66} \textcolor{my_c2}{($\downarrow$ 2.46)}\\ 

    \midrule[1pt]

    &\multicolumn{10}{c}{\textbf{RoBERTa-large}} \\ % 
    \midrule[1pt]

    \rowcolor{gray!10}
    \textbf{Lower Bound}& No-FT$^{\dagger}$ & 49.95 & 49.95 & 25.24 & 25.24 & 9.98 & 9.98 & 5.02 & 5.02 & / \\

    \midrule
    
    \multirow{3}{*}{\textbf{\shortstack{Memory\\Unaware}}}& Linear Probing~\cite{kornblith2019better} & 66.45 & 62.24 & 69.88 & 62.67 & 50.35 & 48.03 & 60.93 & 51.64 & 59.02 \textcolor{my_c2}{($\downarrow$ 22.29)} \\
    
    &FedAdapter~\cite{cai2022fedadapter}                  
                            & --- & --- & 78.59 & 73.87 & --- & --- & --- & --- & 35.30 \textcolor{my_c2}{($\downarrow$ 46.01)}  \\ 
    & C2A~\cite{kim2023client}                       
                            & --- & --- &  76.93 & 71.98  & --- & --- & --- & --- & 34.85 \textcolor{my_c2}{($\downarrow$ 46.46)} \\ 
                            
    \midrule
    
    \multirow{5}{*}{\textbf{\shortstack{Memory\\Aware}}} 
    & FwdLLM~\cite{xu2023fwdllm}                      
                            & 76.89 & 72.46 & 80.41 & 74.52 & 58.97 & 55.24 & 5.13$^*$ & 5.05$^*$ & 53.58 \textcolor{my_c2}{($\downarrow$ 27.73)} \\
    &FedKSeed~\cite{qin2023federated}
    					& 78.19 & 73.76 & 80.63 & 79.85 & 60.12 & 56.75 & 67.80 & 62.93 &70.00 \textcolor{my_c2}{($\downarrow$ 11.31)} \\
    &FLoRA~\cite{wang2024flora}
    						& --- & --- & 80.73 & 79.88 & --- & --- & --- & --- & 36.31 \textcolor{my_c2}{($\downarrow$ 45.00)} \\
    &FedRA~\cite{su2024fedra}                       
                            & 77.45 & 73.71  &  81.56 & 76.63  & 61.69 & 58.37 & 69.27 & 65.31&70.50 \textcolor{my_c2}{($\downarrow$ 10.81)}\\
    &\our\cellcolor{my_c1!50}       &  \textbf{86.47}\cellcolor{my_c1!50} & \textbf{83.08}\cellcolor{my_c1!50} & \textbf{91.96}\cellcolor{my_c1!50} & \textbf{89.52}\cellcolor{my_c1!50} & \textbf{72.77}\cellcolor{my_c1!50} & \textbf{69.24}\cellcolor{my_c1!50} & \textbf{81.12}\cellcolor{my_c1!50} & \textbf{76.32}\cellcolor{my_c1!50} & \textbf{81.31}\cellcolor{my_c1!50} \\ 
    
    \midrule
    \textbf{Upper Bound} & Full Adapters$^{\dagger}$ & \underline{84.93} & \underline{82.71} & \underline{90.13} & \underline{87.06} & \underline{70.94} & \underline{67.53} & \underline{78.76} & \underline{75.57} & \underline{79.70} \textcolor{my_c2}{($\downarrow$ 1.61)}\\
    \bottomrule[1pt]
    \end{tabular}
}
\caption{Performance comparison of \our versus baselines on text classification tasks. \textbf{Bold} and \underline{underlined} values indicate the best and second-best results, respectively. "$-$" denotes methods that failed due to memory constraints (defaulting to No-FT$^{\dagger}$ performance), while "$*$" indicates non-convergence.}
\label{tab_overall_performance}
\vspace{-4mm}
\end{table*}

\subsection{Baselines}

We evaluate \our against a comprehensive suite of baselines, including a performance lower bound (No-FT$^{\dagger}$, denoting the pre-trained model) and an idealized upper bound (Full Adapters$^{\dagger}$, representing memory-unconstrained end-to-end training). We further compare against two categories of state-of-the-art approaches: 1) Memory-Unaware Methods: Linear Probing~\cite{kornblith2019better}, FedAdapter~\cite{cai2022fedadapter}, and C2A~\cite{kim2023client}; and 2) Memory-Aware Methods: FwdLLM~\cite{xu2023fwdllm}, FedKSeed~\cite{qin2023federated}, FLoRA~\cite{wang2024flora}, and FedRA~\cite{su2024fedra}. For detailed descriptions of these methods, please refer to Appendix~\ref{appendix_baseline}.

\subsection{Overall Evaluation}

% Table~\ref{tab_overall_performance} presents a comprehensive performance comparison between \our and baselines. The results show that \our consistently achieves superior performance across all experimental settings, with average accuracy improvements of up to 27.73\%.

Table~\ref{tab_overall_performance} summarizes the main results, showing that \our consistently outperforms baselines across all experimental settings, achieving average accuracy improvements of up to 46.46\%.

% \noindent$\bullet$\textbf{Comparison with Memory-Unaware Methods.} These methods experience performance degradation as model complexity increases.
% This decline primarily results from the growing memory demands of fine-tuning larger models, which limits the participation of resource-constrained devices. For example, FedAdapter exhibits average performance drops of 5.42\% and 9.56\% on DistilBERT and BERT, respectively, and fails to work on RoBERTa across multiple datasets due to memory constraints. 
% C2A exhibits a similar pattern of performance degradation, as it also fails to effectively leverage the valuable data from memory-constrained devices. Compared to \our, it yields average performance drops of 7.35\% on DistilBERT and 11.74\% on BERT. On RoBERTa, C2A remains non-functional on most benchmarks due to excessive memory requirements.
% Additionally, we observe that Linear Probing consistently underperforms compared to FedAdapter and C2A in most scenarios, demonstrating that only fine-tuning the output layer is insufficient for adapting to downstream tasks.

\noindent$\bullet$ \textbf{Comparison with Memory-Unaware Methods.} The fundamental limitation of these methods is their inability to scale. As model complexity increases (DistilBERT $\to$ RoBERTa), performance degrades sharply due to the memory wall. This barrier excludes resource-constrained devices, causing a catastrophic loss of data diversity that compromises the global model. Moreover, FedAdapter and C2A collapse on RoBERTa, failing to run on most datasets due to memory constraints. Even on smaller models where they are functional, their performance lags significantly behind \our; on BERT, they are outperformed by 9.56\% and 11.74\%, respectively. Furthermore, the consistently poor performance of Linear Probing confirms that merely fine-tuning the output layer is insufficient for effective task adaptation.

\noindent$\bullet$ \textbf{Comparison with Memory-Aware Methods.} FwdLLM and FedKSeed reduce activation memory but ignore the dominant parameter bottleneck. Furthermore, their reliance on imprecise gradient estimation severely hurts performance. Compared to \our, FwdLLM suffers average drops of 9.68\%, 23.89\%, and 27.73\% on DistilBERT, BERT, and RoBERTa, respectively, and fails to converge on 20NEWS. Similarly, FedKSeed lags by 7.44\%–11.31\%. FLoRA exhibits degradation patterns similar to memory-unaware methods; since LoRA modules constitute a negligible fraction of memory, reducing their rank offers minimal relief. Although FedRA addresses memory limitations, its random allocation of parameter update tasks leads to synchronization issues, resulting in a significant average performance degradation of up to 10.81\% on RoBERTa compared to \our.

\noindent$\bullet$ \textbf{Comparison with Upper Bound.}
\our even outperforms Full Adapters$^\dagger$ across all evaluation settings, achieving average performance gains of 2.86\%, 2.46\%, and 1.61\% on DistilBERT, BERT, and RoBERTa, respectively, while reducing peak memory usage by up to 16.87$\times$ on RoBERTa.
This superior performance is attributed to its efficient parameter updating strategy, which selectively fine-tunes task-critical layers while minimizing interference with other layers, thereby enhancing task adaptation and knowledge retention.

\begin{figure}[!t]
    \centering
    \begin{subfigure}{0.235\textwidth}
        \centering
        \includegraphics[width=\linewidth]{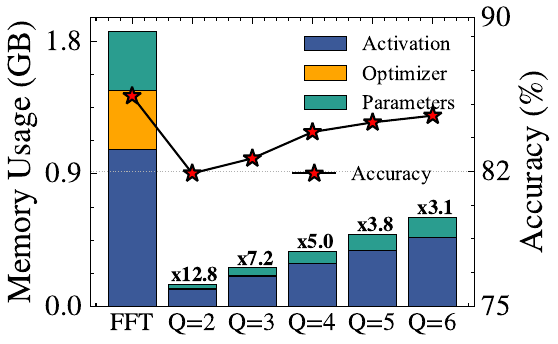}
        \caption{YELP-P (non-IID).}
        \label{Co_tuning1}
    \end{subfigure}
    \hfill
    \begin{subfigure}{0.235\textwidth}
        \centering
        \includegraphics[width=\linewidth]{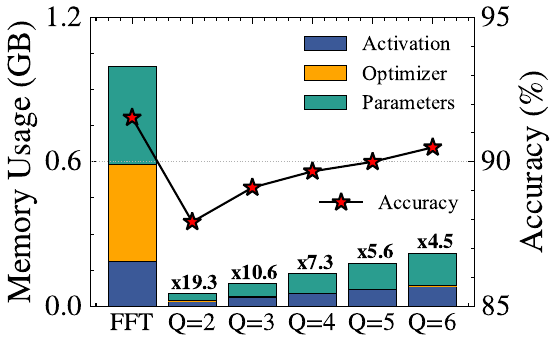}
        \caption{AGNEWS (non-IID).}
        \label{Co_tuning2}
    \end{subfigure}
    \caption{Impact of the co-tuning window size ($Q$) on model performance and memory usage for BERT.}
    \label{Co_tuning}
    \vspace{-4mm}
\end{figure}

% \subsection{Analysis of the Co-Tuning Window Size}

% We now analyze the impact of the co-tuning window size ($Q$), which controls the trade-off between model performance and memory usage. We conduct this analysis on BERT, varying $Q$ from 2 to 6. As shown in Figure~\ref{Co_tuning}, increasing $Q$ consistently improves model performance but also raises peak memory usage. This is an expected trade-off: a larger window facilitates stronger parameter co-adaptation across more layers, but at the cost of loading more components into memory simultaneously. This finding informs our strategy for setting $Q$ in practice: it is determined by the device with the minimum memory capacity, ensuring no device is excluded from training. Notably, with a window size of just $Q=6$, \our matches the performance of full-parameter fine-tuning (FFT), while reducing peak memory usage by up to 4.5$\times$, further highlighting the effectiveness of the co-tuning mechanism.

\subsection{Analysis of the Co-Tuning Window Size}

We now analyze the impact of the co-tuning window size ($Q$), which controls the trade-off between model performance and memory usage. We conduct this analysis on BERT, varying $Q$ from 2 to 6. As illustrated in Figure~\ref{Co_tuning}, increasing $Q$ yields consistent performance gains but proportionally raises peak memory consumption. This reflects a fundamental trade-off: a larger window facilitates stronger parameter co-adaptation across more layers, but at the cost of loading more components into memory simultaneously. In practice, we set $Q$ based on the device with the lowest capacity to ensure inclusive participation. Notably, with $Q=6$, \our matches the performance of full-parameter fine-tuning (FFT), while reducing peak memory usage by up to 4.5$\times$, highlighting the effectiveness of the co-tuning mechanism.

% In this section, we investigate the effect of varying the number of co-tuning adapters, denoted by $Q$, on model performance and memory consumption. Specifically, we conduct experiments on BERT with $Q \in \{2, 3, 4, 5, 6\}$. Figure~\ref{Co_tuning} illustrates the experimental results. We observe that as $Q$ increases, model performance consistently improves, accompanied by a rise in memory usage. This is because co-tuning more adapters facilitates stronger parameter co-adaptation across layers, but also requires more layers to be simultaneously loaded into memory. Therefore, to balance memory efficiency and model performance, we set $Q$ according to the minimum memory capacity across participating devices, ensuring that all devices can participate without exclusion. Moreover, when $Q=6$, \our matches the performance of full fine-tuning (FFT), while reducing peak memory usage by up to 4.5$\times$, further highlighting the effectiveness of the co-tuning mechanism.

% \input{Table/figure_lambda}

\subsection{Analysis of the Global Loss Weight}

In this section, we explore the impact of the global loss weight ($\lambda$), which balances the local and global learning objectives. We conduct experiments on DistilBERT, varying $\lambda$ across the set \{0.0,0.1,0.2,0.5,1.0\}. 
Figure~\ref{fig:training_size} shows that setting $\lambda=0$ yields the lowest performance, confirming that purely local optimization induces myopic learning. Conversely, integrating the global signal boosts accuracy by up to 22.76\%. However, an excessively high $\lambda$ (e.g., 1.0) degrades performance by 10.27\% (compared to $\lambda=0.2$), as the global objective overshadows layer-specific feature learning.
Therefore, we recommend setting $\lambda=0.1$ for BERT and 0.2 for other models in practice.

% a complete absence of the global objective ($\lambda=0$) yields the weakest performance, confirming that a purely local focus leads to myopic learning. Integrating the global loss signal results in a substantial performance gain of up to 22.76\%. However, an excessively high $\lambda$ also degrades performance. For instance, at $\lambda=1.0$, accuracy drops by up to 10.27\% (compared to $\lambda=0.2$), as the global objective overshadows the learning of critical, layer-specific features. 

\definecolor{red}{RGB}{172,21,28}
\definecolor{blue}{RGB}{39,89,167}
\definecolor{red1}{RGB}{203,104,104}
\definecolor{blue1}{RGB}{104,155,203}
\definecolor{color1}{HTML}{283c63}
\definecolor{color2}{HTML}{00ad7c}

\begin{figure}[!t]
\centering
\hspace{-6mm}
\begin{tikzpicture}
    \scriptsize{
    \begin{axis}
    [
        anchor=north west,
        at={(-0em,-5em)},
        ymajorgrids,
        xmajorgrids,
        grid style=dashed,
        width=.25\textwidth,
        height=.20\textwidth,
        yticklabel style={/pgf/number format/precision=0,/pgf/number format/fixed zerofill,scale=1.0},
        xmax=2100,
        xmin=300,
        ymin=55,
        ymax=95,
        xtick={400,800,1200,1600,2000},
        xticklabels={0.0,0.1,0.2,0.5,1.0},
        ytick={55,65,75,85,95},
        xlabel={\scriptsize{(a) Different $\lambda$ (IID).}},
        xlabel style={scale=1.2, yshift=0.2em, xshift=0.1em},
        ylabel=\footnotesize{\scriptsize Accuracy (\%)},
        ylabel style={yshift=0.0em, scale=1.2},
        legend style={at={(2.4,1.4)}, anchor=north east, font={\tiny}, cells={anchor=west}, fill opacity=0.8, scale=1.0, legend columns=3}
        ]

        \addplot[red,mark=pentagon*,,mark size=2.5pt,thick,mark options={fill=white,draw=red,line width=1pt}] coordinates {(400,65.84) (800,74.08) (1200,85.08) (1600,81.46) (2000,76.83)};
        \addlegendentry{\scalebox{1.2}{YELP-P}}

        \addplot[color1,mark=*,mark size=2.5pt,thick,mark options={fill=white,draw=color1,line width=1pt}] coordinates {(400,83.25) (800,86.48) (1200,91.29) (1600,87.57) (2000,85.04)};
        \addlegendentry{\scalebox{1.2}{AGNEWS}}

        \addplot[color2,mark=diamond*,mark size=3pt,thick,mark options={fill=white,draw=color2,line width=1pt}] coordinates {(400,61.72) (800,66.61) (1200,72.54) (1600,70.47) (2000,68.38)};
        \addlegendentry{\scalebox{1.2}{YAHOO}} 
    \end{axis}
	
  \begin{axis}
    [
        anchor=north west,
        at={(16em,-5em)},
        ymajorgrids,
        xmajorgrids,
        grid style=dashed,
        width=.25\textwidth,
        height=.20\textwidth,
        yticklabel style={/pgf/number format/precision=0,/pgf/number format/fixed zerofill,scale=1.0},
        xmax=2100,
        xmin=300,
        ymin=50,
        ymax=92,
        xtick={400,800,1200,1600,2000},
        xticklabels={0.0,0.1,0.2,0.5,1.0},
        ytick={50,60,70,80,90},
        xlabel={\scriptsize{(b) Different $\lambda$ (non-IID).}},
        xlabel style={scale=1.2, yshift=0.2em, xshift=0.1em},
        ylabel=\footnotesize{\scriptsize Accuracy (\%)},
        ylabel style={yshift=0.em, scale=1.2},
        legend style={at={(1.8,1.2)}, anchor=north east, font={\tiny}, cells={anchor=west}, fill opacity=0.8, scale=1.0, legend columns=3}
        ]

        \addplot[red,mark=pentagon*,,mark size=2.5pt,thick,mark options={fill=white,draw=red,line width=1pt}] coordinates {(400,60.13) (800,70.76) (1200,82.89) (1600,77.98) (2000,72.62)};
        % \addlegendentry{\scalebox{1.2}{YELP-P}}

        \addplot[color1,mark=*,mark size=2.5pt,thick,mark options={fill=white,draw=color1,line width=1pt}] coordinates {(400,79.47) (800,83.75) (1200,88.05) (1600,84.05) (2000,81.25)};
        % \addlegendentry{\scalebox{1.2}{AGNEWS}}

        \addplot[color2,mark=diamond*,mark size=3pt,thick,mark options={fill=white,draw=color2,line width=1pt}] coordinates {(400,56.70) (800,62.35) (1200,69.18) (1600,67.03) (2000,64.68)};
        % \addlegendentry{\scalebox{1.2}{YAHOO}} 
    \end{axis}}   
\end{tikzpicture}
    \vspace{-2mm}
    \caption{Impact of the global loss weight on model performance for DistilBERT.}
    \label{fig:training_size}
    \vspace{-4mm}
\end{figure}
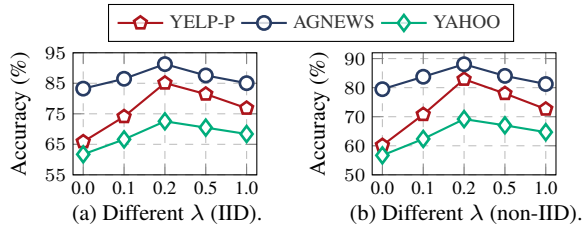

% we explore the impact of the globally perceptive optimization strategy on model performance. Specifically, we conduct experiments on DistilBERT and vary the influence of the global loss by adjusting the value of $\lambda$ ($\lambda\in\{0.0, 0.1, 0.2, 0.5, 1.0\}$). Figure~\ref{fig:training_size} presents the experimental results, showing that integrating the global loss ($\lambda \neq 0$) consistently improves model performance compared to the case where $\lambda = 0.0$, with a performance gain of up to 22.76\%. These results suggest that focusing exclusively on the local objective can lead to significant information loss, ultimately compromising model performance.
% GPO addresses this limitation by aligning adapter updates with the model's overall objective, thereby promoting coordinated learning across layers.
% Additionally, we also observe that an excessively high $\lambda$ value still degrades model performance. For instance, when $\lambda = 1.0$, performance drops by up to 10.27\% due to the overemphasis on the global objective, which hinders the local optimization process from effectively capturing critical features.

\definecolor{color1}{HTML}{f2e9d0}

\begin{table}[!t]
\small
\centering
\resizebox{1\linewidth}{!}{
\begin{tabular}{lcccc}
\toprule[1pt]
\textbf{Threshold}& \textbf{IID} & \textbf{non-IID}  &\multirow{1}{*}{\textbf{Speedup}} &\multirow{1}{*}{\textbf{Comm. Reduction}} \\  
\midrule[1pt]

\multicolumn{5}{c}{\textbf{DistilBERT-base (YELP-P)}}\\

\midrule[1pt]

% \rowcolor{color1!20}

%Full Fine-tuning$^\dagger$ & 87.39 & 85.05  &$\times$1&$\times$1 \\ 
Full Adapters$^\dagger$ & 84.76 & 82.58 & $\times$1 & $\times$1 \\ % $\times$1.23 $\times$142.23
\midrule
$T = 1.0$ & 82.76 & 80.03 & $\times$1.72 & $\times$1.66 \\ % $\times$2.12 $\times$235.76

$T = 0.9$ & 84.43 & 82.37 & $\times$1.86 & $\times$2.49 \\ % $\times$2.29 \times$353.64
$T = 0.8$ & \textbf{86.53} & \textbf{84.13} & \textbf{$\times$1.97} & \textbf{$\times$2.75} \\ % \textbf{$\times$2.42} \textbf{$\times$391.52}
%Full Adapters$^\dagger$ & 84.76 & 82.58 & $\times$1.23 & $\times$142.23 \\

\midrule[1pt]
\multicolumn{5}{c}{\textbf{DistilBERT-base (AGNEWS)}}\\ 
\midrule[1pt]
% \rowcolor{color1!20}

%Full Fine-tuning$^\dagger$ & 93.41 & 91.13&$\times$1&$\times$1 \\ 
Full Adapters$^\dagger$ &  90.05 & 86.96  & $\times$1 & $\times$1 \\ % $\times$1.24 $\times$90.51 
\midrule
$T = 1.0$ & 88.45 & 85.82 & $\times$1.78 & $\times$2.66 \\ % $\times$2.21 $\times$240.72

$T = 0.9$ & 90.24 & 87.86 & $\times$1.94 & $\times$3.13 \\ % $\times$2.40 $\times$282.91
$T = 0.8$ & \textbf{92.59} & \textbf{90.25} & \textbf{$\times$2.02} & \textbf{$\times$3.47} \\ % \textbf{$\times$2.51} \textbf{$\times$314.35}
%Full Adapters$^\dagger$ &  90.05 & 86.96  & $\times$1.24 & $\times$90.51 \\ 

\bottomrule[1pt]
\end{tabular}
}
\caption{Impact of the fine-tuning threshold $T$ on model performance. In this set of experiments, we set $Q=3$.}
\label{tab:impact of threshold}
\vspace{-3mm}
\end{table}

\subsection{Analysis of the Fine-Tuning Threshold}

We then analyze the impact of the fine-tuning threshold ($T$), which determines the starting layer for our chain optimization. Our experiments vary $T$ across the set \{1.0,0.9,0.8\}, with $T=1.0$ representing the baseline case of fine-tuning all layers. 
The results in Table~\ref{tab:impact of threshold} reveal a crucial insight: fine-tuning all layers is suboptimal. Performance peaks at $T=0.8$, surpassing the full-chain baseline ($T=1.0$) by up to 4.1\% on YELP-P and 4.43\% on AGNEWS. This suggests that freezing general-purpose lower layers not only conserves resources but also enhances model generalization. Remarkably, \our ($T=0.8$) outperforms even the idealized Full Adapters$^{\dagger}$, achieving 3.29\% higher accuracy on AGNEWS with 2.02$\times$ faster convergence and 3.47$\times$ lower communication overhead. These results powerfully demonstrate the effectiveness of our Function-Oriented Adaptive Tuning scheme.

\definecolor{Maroon}{HTML}{AE3135}
\definecolor{BLUE}{HTML}{6466AE}
\definecolor{my-green}{HTML}{F5B482}
\definecolor{my-yellow}{HTML}{FFBE7A}
\definecolor{my-blue}{HTML}{82B0D2}

\begin{table}[!t]
\centering
\resizebox{\linewidth}{!}{
\begin{tabular}{lcccccc}
\toprule[1pt]
\textbf{Method}   & \textbf{MMLU} & \textbf{BBH} & \textbf{DROP}&\textbf{CRASS} & \textbf{Average} &\textbf{Mem. Reduction} \\ 
\midrule[1pt]
\multicolumn{7}{c}{\textbf{LLaMA2-7B}} \\ 
\midrule[1pt]
% \rowcolor{gray!20}
Full Adapters$^\dagger$ & 43.68 & 31.72 & 32.19 & 44.95 & 38.14 &$\times$1\\
\midrule

$Q=6$ &  45.17 & 32.36 & 33.40 & 50.63 & 40.39 & \textbf{$\times$4.29}\\
% \rowcolor{cyan!10}
$Q=7$ & 46.04 & 33.19 & 34.05 & 52.87 & 41.54 & $\times$3.69 \\
% \rowcolor{cyan!10}
$Q=8$ & \textbf{46.76} & \textbf{33.65} & \textbf{34.79} & \textbf{55.34} & \textbf{42.64} & $\times$3.23 \\

\midrule[1pt]
\multicolumn{7}{c}{\textbf{LLaMA3.1-8B}} \\ 
\midrule[1pt]

% \rowcolor{mylightred!10}
% \rowcolor{gray!20}
Full Adapters$^\dagger$ & 61.15 & 61.28 & 55.72 & 74.63 & 63.20 &$\times$1\\
\midrule

% \rowcolor{cyan!10}

$Q=6$ &  65.46 & 65.87 & 60.36 & 85.48 & 69.29 & \textbf{$\times$4.60}\\
% \rowcolor{cyan!10}
$Q=7$ & 67.19 & 68.60 & 62.68 & 90.07 & 72.14 & $\times$3.94 \\
% \rowcolor{cyan!10}
$Q=8$ & \textbf{68.01} & \textbf{70.56} & \textbf{63.42} & \textbf{93.64} & \textbf{73.91} & $\times$3.45 \\

\bottomrule[1pt]
\end{tabular}
}
\caption{{Performance of \our on instruction tuning tasks with varying values of $Q$}.}
\label{tab:instruction_tuning}
\vspace{-4mm}
\end{table}

\subsection{Performance on Instruction Tuning}

We further evaluate \our on instruction tuning tasks using LLaMA2-7B and LLaMA3.1-8B, benchmarking against Full Adapters$^{\dagger}$. As detailed in Table~\ref{tab:instruction_tuning}, \our consistently outperforms the baseline while substantially reducing memory usage. 
For instance, on LLaMA2-7B, when \( Q = 6 \), \our yields improvements of 1.49\%, 0.64\%, 1.21\%, and 5.68\% on MMLU, BBH, DROP, and CRASS, respectively, while reducing peak memory usage by 4.29$\times$. When \( Q = 8 \), it achieves a 4.50\% average performance gain with a 3.23$\times$ memory reduction. These benefits are even more pronounced on LLaMA3.1-8B, where average accuracy improves by up to 10.71\% alongside a 3.45$\times$ reduction in memory. This consistently superior performance, achieved with low memory usage, validates \our as a versatile and practical solution for on-device federated fine-tuning.

% We further evaluate the effectiveness of \our on instruction tuning tasks using LLaMA2-7B and LLaMA3.1-8B.
% For benchmarking, Full Adapters$^{\dagger}$ is employed as the baseline. 
% Table~\ref{tab:instruction_tuning} shows that \our consistently achieves higher accuracy while significantly reducing peak memory usage.
% For instance, on LLaMA2-7B, when \( Q = 6 \), \our yields improvements of 1.49\%, 0.64\%, 1.21\%, and 5.68\% on MMLU, BBH, DROP, and CRASS, respectively, while reducing peak memory usage by 4.29$\times$. When \( Q = 8 \), it achieves a 4.50\% average performance gain with a 3.23$\times$ memory reduction.
% These improvements are even more pronounced on LLaMA3.1-8B, with an average performance improvement of up to 10.71\% alongside a 3.45$\times$ memory reduction.
% This consistently superior performance, achieved with low memory usage, validates \our as a versatile and practical solution for on-device federated fine-tuning.

\definecolor{my_c1}{HTML}{c6f1e7}
% \definecolor{my_c2}{HTML}{e4eddb}
\definecolor{my_c2}{HTML}{13829b}

\begin{table}[!t]
\small
\centering
\resizebox{\linewidth}{!}{
\begin{tabular}{lccccc}
\toprule[1pt]
\multirow{2}{*}{\textbf{Method}}  & \multicolumn{2}{c}{\textbf{YELP-P}} & \multicolumn{2}{c}{\textbf{AGNEWS}} &\multirow{2}{*}{\textbf{Average}} \\  

\cmidrule{2-3}
\cmidrule{4-5}

& \textbf{IID} & \textbf{non-IID} & \textbf{IID} & \textbf{non-IID}   \\ 
\midrule[1pt]
\multicolumn{6}{c}{\textbf{DistilBERT-base}} \\ 
\midrule[1pt]

w/o DLCT & 73.63 & 69.04 & 83.89 & 80.45 & 76.75 \textcolor{my_c2}{($\downarrow$ 11.64)}\\ 
w/o GPO  & 70.12 & 63.98 & 83.98 & 80.57 & 74.66 \textcolor{my_c2}{($\downarrow$ 13.73)}\\ 
w/o FOAT & 79.95 & 77.14 & 86.78 & 83.59 & 81.87 \textcolor{my_c2}{($\downarrow$ 6.52)}\\ 
\rowcolor{my_c1!50}
\our &\textbf{86.57} & \textbf{84.22} & \textbf{92.57} & \textbf{90.18} & \textbf{88.39}\\

\midrule[1pt]
\multicolumn{6}{c}{\textbf{BERT-base}} \\ 
\midrule[1pt]

w/o DLCT & 72.83 & 68.27 & 83.34 & 80.13 & 76.14 \textcolor{my_c2}{($\downarrow$ 11.86)}\\ 
w/o GPO  & 69.27 & 63.86 & 84.21 & 80.58 & 74.48 \textcolor{my_c2}{($\downarrow$ 13.52)}\\ 
w/o FOAT & 78.82 & 77.03 & 86.37 & 81.07 & 80.82 \textcolor{my_c2}{($\downarrow$ 7.18)}\\ 
\rowcolor{my_c1!50}
\our & \textbf{86.40}&\textbf{84.05}&\textbf{91.89}&\textbf{89.67}&\textbf{88.00}\\

\midrule[1pt]
\multicolumn{6}{c}{\textbf{RoBERTa-large}} \\ 
\midrule[1pt]

w/o DLCT & 72.12 & 67.98 & 82.50 & 79.96 & 75.64 \textcolor{my_c2}{($\downarrow$ 12.12)}\\ 
w/o GPO  & 72.31 & 67.63 & 85.22 & 81.65 & 76.70 \textcolor{my_c2}{($\downarrow$ 11.06)}\\ 
w/o FOAT & 80.21 & 77.73 & 87.39 & 83.82 & 82.29 \textcolor{my_c2}{($\downarrow$ 5.47)}\\ 
\rowcolor{my_c1!50}
\our &\textbf{86.47} & \textbf{83.08} & \textbf{91.96} & \textbf{89.52} & \textbf{87.76}\\

\bottomrule[1pt]
\end{tabular}
}
\caption{Ablation study of \our. }
\label{ablation_study}
\vspace{-3mm}
\end{table}

\subsection{Ablation Study}

Finally, we conduct an ablation study to quantify the individual contributions of \our's core components: DLCT, GPO, and FOAT. The results (Table~\ref{ablation_study}) confirm that removing any technique precipitates a substantial performance drop, validating their collective necessity. For DistilBERT, eliminating DLCT, GPO, or FOAT reduces average accuracy by 11.64\%, 13.73\%, and 6.52\%, respectively. This impact persists as model complexity scales: on RoBERTa, performance degrades by 12.12\% (w/o DLCT), 11.06\% (w/o GPO), and 5.47\% (w/o FOAT). These findings underscore that each mechanism plays an indispensable role in ensuring the effectiveness of chain optimization.

% Finally, we conduct an ablation study to isolate and quantify the contribution of each core component in \our, i.e., DLCT, GPO, and FOAT. 
% The results, presented in Table~\ref{ablation_study}, show that removing any technique leads to a substantial drop in performance, confirming that all are integral to \our's success. 
% For instance, on DistilBERT, the average accuracy decreases by 11.64\% without DLCT, 13.73\% without GPO, and 6.52\% without FOAT. As the global model complexity increases, the contribution of each technique remains pronounced. Specifically, for RoBERTa, the average accuracy drops by 12.12\% without DLCT, 11.06\% without GPO, and 5.47\% without FOAT.
% These results demonstrate that these techniques play a significant role in guiding the chain optimization process.
\section{Conclusion}

In this paper, we propose \our, an innovative federated fine-tuning paradigm that effectively addresses the memory constraints of participating devices via chain optimization.
\our further integrates three novel techniques: 1) Dynamic Layer Co-Tuning, 2) Globally Perceptive Optimization, and 3) Function-Oriented Adaptive Tuning. Extensive experiments on multiple benchmarks demonstrate that \our significantly outperforms state-of-the-art baselines in model performance and resource efficiency.
\section*{Limitations}
Despite the promising results and memory efficiency demonstrated by \our, we acknowledge several limitations that present avenues for future research. First, our evaluation is currently confined to natural language processing tasks using Transformer-based architectures (e.g., BERT and LLaMA series). While \our is theoretically applicable to other neural networks, its effectiveness on distinct architectures (such as SSMs or Mamba) and other modalities (e.g., Multimodal LLMs) remains to be empirically verified. Second, while \our inherently preserves data privacy by keeping raw data local, we have not yet integrated it with advanced privacy-enhancing technologies, such as Differential Privacy (DP) or Homomorphic Encryption. Exploring the trade-off between the noise introduced by DP and the precise feature alignment required by our layer-wise tuning remains a valuable direction for future work.

\section*{Ethical Considerations}

% Our research proposes \our to enable privacy-preserving LLM fine-tuning on edge devices. This work aligns with ethical guidelines by promoting data privacy; our method ensures that sensitive user data remains on local devices, addressing key privacy concerns in centralized AI training. Furthermore, by drastically reducing memory requirements, \our promotes sustainability and inclusivity, enabling the deployment of LLMs on consumer-grade hardware rather than relying solely on energy-intensive data centers. All datasets used in this study are public benchmarks. We acknowledge the general risks associated with LLM generation (e.g., hallucination or toxicity) and emphasize that standard safety alignment procedures should be applied alongside our optimization framework when deploying in production environments.

Our research proposes \our to enable privacy-preserving LLM fine-tuning on edge devices. This work aligns with ethical guidelines by promoting data privacy; our method ensures that sensitive user data remains on local devices, addressing key privacy concerns in centralized AI training. Furthermore, by drastically reducing memory requirements, \our promotes sustainability and inclusivity, enabling the deployment of LLMs on consumer-grade hardware rather than relying solely on energy-intensive data centers. 
All datasets and models used in this study are public benchmarks and pre-trained models (e.g., BERT, RoBERTa, LLaMA2, LLaMA3.1) that are utilized strictly in accordance with their intended research purposes and licensing terms. We strictly adhere to all applicable licenses and usage policies. All evaluation benchmarks are employed exclusively for assessing model capabilities, and all pre-trained models are used solely for research evaluation purposes consistent with their intended use cases. 

We acknowledge the general risks associated with LLM generation (e.g., hallucination or toxicity) and emphasize that standard safety alignment procedures should be applied alongside our optimization framework when deploying in production environments.

\section*{Acknowledgements}

This work is supported in part by the Science and Technology Development Fund of Macau (0107/2024/RIA2, 0061/2025/RIB2), Joint Science and Technology Research Project with Hong Kong and Macau in Key Areas of Nansha District's Science and Technology Plan (EF2024-00180-IOTSC) and the Multi-Year Research Grant of University of Macau (MYRG-GRG2023-00211-IOTSC-UMDF, MYRG-GRG2024-00180-IOTSC).
\bibliography{custom}

% \newpage
\appendix
% \onecolumn
\section{The Definition of HSIC}\label{appendix_hsic}
The HSIC is computed as:
\begingroup
\small  % <--- 这里控制大小，可选 \small, \footnotesize, \scriptsize, \tiny
\begin{align}
\text{HSIC}(Z_{i}, Z_{j}) 
&= \mathbb{E}_{Z_{i},Z_{i}',Z_{j},Z_{j}'}\left[ k_{Z_{i}}(Z_{i}, Z_{i}')k_{Z_{j}}(Z_{j}, Z_{j}') \right] \nonumber \\
&\quad + \mathbb{E}_{Z_{i},Z_{i}'}\left[ k_{Z_{i}}(Z_{i}, Z_{i}') \right]
        \mathbb{E}_{Z_{j},Z_{j}'}\left[ k_{Z_{j}}(Z_{j}, Z_{j}') \right] \nonumber \\
&\quad - 2\mathbb{E}_{Z_{i},Z_{j}}\left[
        \mathbb{E}_{Z_{i}'}[k_{Z_{i}}(Z_{i}, Z_{i}')] 
        \mathbb{E}_{Z_{j}'}[k_{Z_{j}}(Z_{j}, Z_{j}')]
    \right],
\label{eq:hsic}
\end{align}
\endgroup
where $k_{Z_{i}}$ and $k_{Z_{j}}$ denote kernel functions. 

\section{The Workflow of \our}\label{appendix_algo}

Algorithm~\ref{alg:profit} presents the complete workflow of \our.

\begin{algorithm}[t]
\caption{The workflow of \our}
\label{alg:profit}
\begin{algorithmic}[1]
\Require 
$N$ devices with memory budgets $\{\mathrm{Mem}_n\}_{n=1}^N$; 
$R$ communication rounds; 
Hyperparameters $\lambda$ and $T$; 
Global model $\Theta$.
\Ensure Fine-tuned global model $\Theta^*$.

\Statex \hspace{-\algorithmicindent}\colorbox{algo_col!50}{\textit{Phase 1: Pre-Training Setup}}
\State \textbf{All devices}: Compute local CKA scores on $\Theta$ via Eq.~\eqref{eq_CKA}.
\State \textbf{Server}: Aggregates similarity scores and sets $L_{start}$ based on threshold $T$. \Comment{FOAT}
\State \textbf{Server}: Sets the co-tuning window size $Q$ based on the minimum device memory, $min(\{\mathrm{Mem}_n\}_{n=1}^N)$. \Comment{DLCT}

\Statex \hspace{-\algorithmicindent}\colorbox{algo_col!50}{\textit{Phase 2: Iterative Federated Fine-Tuning}}

\For{round $r = 1$ to $R$}
    \State Server samples a subset of devices $\mathcal{S}_r$.
    % \State Server determines the current training window of adapters, e.g., $[\text{Adapter}_s, ..., \text{Adapter}_{s+Q-1}]$. \Comment{DLCT}
    \State Server sends relevant parameters to $\mathcal{S}_r$. \Comment{DLCT}
    \ForAll{device $n \in \mathcal{S}_r$ \textbf{in parallel}}
        \State Perform local training using Eq.~\eqref{update_objective}. \Comment{GPO}
        \State Upload local adapter updates $\Delta\Theta_{n}^{r}$.
    \EndFor
    \State Server aggregates local updates to refine the global model:
    $\Theta_{r+1} \leftarrow \Theta_{r} + \sum_{n \in \mathcal{S}_r} \frac{|D_{n}|}{\sum_{n \in \mathcal{S}_r}|D_{n}|} \Delta\Theta_{n}^{r}$.
\EndFor
\State \Return $\Theta^* \leftarrow \Theta$.
\end{algorithmic}
\end{algorithm}

\section{Theoretical Analysis}\label{appendix_convergence}

In this section, we present a rigorous convergence analysis of \our in the context of adapter-based federated fine-tuning. Leveraging standard assumptions and established results from the federated optimization literature~\cite{li2019convergence, wang2022progfed}, we show that \our converges to a stationary point, thus providing formal guarantees for its optimization dynamics.

\subsection{Preliminaries}

In federated fine-tuning, the global optimization objective is defined as:
\begin{equation}
  \min_{\Theta} F(\Theta) = \sum_{n=1}^{N} \frac{|D_{n}|}{|D|} \, \mathcal{L}_{n}(\Theta; D_{n}),
  \label{global_obj}
\end{equation}
where \( \Theta \) denotes the set of global model parameters, \( D_n \) is the local dataset residing on device \( n \) (\( n \in [1, N] \)), and \( |D_n| \) and \( |D| \) represent the sizes of the local and total datasets, respectively. The function \( \mathcal{L}_{n}(\Theta; D_{n}) \) denotes the local optimization objective on device \( n \).

In \our, adapter modules are integrated into a pre-trained LLM and fine-tuned in a chain, layer-wise manner. The adaptation process at each layer adheres to the transformation defined in Equation~\eqref{eq_adapter_trans}. Furthermore, the objective in each stage combines a local loss and a globally aligned loss, computed through an auxiliary output branch, as formulated in Equation~\eqref{update_objective}. This strategy facilitates stable layer-wise adaptation while ensuring alignment with the overall global optimization goal.

\subsection{Assumptions}
For analytical purposes, we adopt a set of assumptions that follow established formulations in FL~\cite{li2019convergence}, while extending them to accommodate the characteristics of adapter-based fine-tuning.

\begin{enumerate}
    \item \textbf{Smoothness.}  
    Each local objective function \( \mathcal{L}_{n}(\Theta; D_{n}) \) is differentiable and has an \( L \)-Lipschitz continuous gradient, meaning there exists a constant \( L > 0 \) such that for any \( \Theta_{1}, \Theta_{2} \), the following holds:
    \begin{equation}
      \small
      \|\nabla \mathcal{L}_{n}(\Theta_{1}; D_{n}) - \nabla \mathcal{L}_{n}(\Theta_{2}; D_{n})\| \le L\, \|\Theta_{1} - \Theta_{2}\|.
    \end{equation}

    This condition guarantees that the gradient of the local objective varies smoothly with respect to the model parameters, enabling stable gradient-based optimization.

    \item \textbf{Unbiased Gradients and Bounded Variance.} 
    The stochastic gradient computed on each device provides an unbiased estimate of the true gradient with bounded variance. Formally, there exist constants $\sigma^{2}$ and $G>0$ such that:
    \begin{equation} 
    \mathbb{E}[\nabla \mathcal{L}_{n}(\Theta; D_n, \xi_n)] = \nabla \mathcal{L}_{n}(\Theta; D_n), 
    \end{equation}
    \begin{equation} 
    \mathbb{E}[\|\nabla \mathcal{L}_{n}(\Theta; D_n, \xi_n) - \nabla \mathcal{L}_{n}(\Theta; D_n)\|^2] \leq \sigma^2
    \end{equation}
    \begin{equation} 
    \|\nabla \mathcal{L}_{n}(\Theta; D_n)\| \leq G, \quad \forall n, \Theta.
    \end{equation}
    This assumption ensures that stochastic optimization does not introduce excessive variance, thereby promoting stable convergence during training.

    \item \textbf{Bounded Heterogeneity.} 
    Data heterogeneity across devices can lead to divergence in local gradients. We assume that the discrepancy between local gradients across devices is bounded. Specifically, there exists a constant \( H > 0 \) such that:
    \begin{equation}
    \small
    \mathbb{E}\left[\|\nabla \mathcal{L}_{m}(\Theta; D_m) - \nabla \mathcal{L}_{n}(\Theta; D_n)\|^2\right] \leq H^2,~ \forall m, n.
    \end{equation}
    This assumption captures the impact of non-IID data in federated settings. A smaller value of \( H \) implies greater homogeneity among devices, which typically leads to faster and more stable convergence.

    \item \textbf{Adapter Error Bound.}
    In \our, fine-tuning is performed on a subset of layers selected by the FOAT mechanism, which introduces an approximation error due to the exclusion of certain layers. We assume this error is bounded by a threshold-dependent constant \( \epsilon_{\text{FOAT}} \ge 0 \). Formally, there exists:
    \begin{equation}
    \left\| \nabla \mathcal{L}_{n}(\Theta_{\text{FOAT}}; D_n) - \nabla \mathcal{L}_{n}(\Theta_{\text{Full}}; D_n) \right\| \leq \epsilon_{\text{FOAT}},
    \end{equation}
    where \( \Theta_{\text{FOAT}} \) denotes the parameters updated in the selected layers, and \( \Theta_{\text{Full}} \) corresponds to full-model fine-tuning. With appropriate layer selection, this error can be made arbitrarily small, ensuring FOAT preserves optimization effectiveness while improving efficiency.

\end{enumerate}

\subsection{Proof of Convergence}
\noindent\textbf{One-Round Descent Lemma.} 
Let \( \Theta_t \) denote the global model parameters at the \( t \)-th iteration, and define the aggregated gradient across participating devices as:
\begin{equation}
  g_t = \sum_{n=1}^{N} \frac{|D_{n}|}{|D|}\, \nabla \mathcal{L}_{n}(\Theta_t; D_{n}).
\end{equation}

In the analysis of FedAvg-based algorithms~\cite{li2019convergence, li2022one, tam2023federated,tian2022harmony}, it is well established that under the smoothness assumption, selecting an appropriate learning rate \( \eta_t \) ensures the following descent property:
\begin{equation}
  F(\Theta_{t+1}) \le F(\Theta_t) - \eta_t\, \|g_t\|^2 + \frac{L\eta_t^2}{2}\, \|g_t\|^2.
  \label{basic_decrease}
\end{equation}

Thus, if \( \eta_t \) is small enough, i.e., \( \eta_t < \frac{2}{L} \), the term \( \frac{L\eta_t^2}{2}\, \|g_t\|^2 \) remains a higher-order error, ensuring a sufficient decrease in \( F(\Theta) \) per iteration.

\noindent\textbf{Roles of Mechanisms in \our.}  
In contrast to FedAvg, \our incorporates two key mechanisms that reshape the optimization dynamics:

\begin{itemize}
    \item \textbf{Globally Perceptive Optimization (GPO).}  
    The GPO strategy customizes the update objective for each layer by incorporating both the standard local loss and an auxiliary global loss, as defined in Equation~\eqref{update_objective}. Let \( g_t^{\text{local}} \) denote the gradient from the local loss, and \( g_t^{\text{global}} \) denote the gradient from the auxiliary global branch. The effective gradient becomes:
    \begin{equation}
      \tilde{g}_t = g_t^{\text{local}} + \lambda\, g_t^{\text{global}}.
    \end{equation}
    The global term \( g_t^{\text{global}} \) serves as a regularization force, aligning local updates with the global objective and mitigating optimization drift across devices. With a properly chosen coefficient \( \lambda \) and well-designed auxiliary branches, the bias introduced by using \( \tilde{g}_t \) instead of the full end-to-end gradient is bounded by a constant \( \epsilon_{\text{aux}} \ge 0 \).

    \item \textbf{Dynamic Layer Co-Tuning (DLCT).} The DLCT mechanism ensures that the loss landscape observed by concurrently fine-tuned adapters is more coherent, effectively reducing inter-layer gradient isolation. Therefore, the effective descent relation becomes:
    \begin{equation}
      F(\Theta_{t+1}) \le F(\Theta_t) - \eta_t \Bigl(\|\tilde{g}_t\|^2 - \epsilon_{\text{aux}} - \epsilon_{\text{FOAT}}\Bigr).
      \label{layered_decrease}
    \end{equation}

\end{itemize}

With appropriate hyperparameter tuning, \( \epsilon_{\text{aux}} \) and \( \epsilon_{\text{FOAT}} \) 
can be made arbitrarily small.
As a result, there exists a learning rate schedule \( \{\eta_t\} \) (or a sufficiently small constant \( \eta \)) such that the sequence \( \{F(\Theta_t)\} \) is monotonically decreasing and lower-bounded, ensuring convergence.
 
\medskip

\noindent\textbf{Theorem 1 (Multi-Round Convergence).}
Assume that the global objective \( F(\Theta_t) \) converges to a finite limit \( F^* \) as \( t \to \infty \). Then, from the descent inequality in Equation~\eqref{layered_decrease}, it follows that:
\begin{equation}
  \lim_{t\to\infty} \eta_t \left(\|\tilde{g}_t\|^2 - \epsilon_{\text{aux}} - \epsilon_{\text{FOAT}}\right) = 0.
\end{equation}

Since the learning rate \( \eta_t \) remains positive and does not decay too rapidly (as ensured by standard stochastic approximation conditions), we have:
\begin{equation}
  \lim_{t\to\infty} \|\tilde{g}_t\|^2 \le \epsilon_{\text{aux}} + \epsilon_{\text{FOAT}}.
\end{equation}

By appropriately tuning hyperparameters, the approximation errors \( \epsilon_{\text{aux}} \) and \( \epsilon_{\text{FOAT}} \) can be made arbitrarily small, such that:
\begin{equation}
  \epsilon_{\text{aux}} + \epsilon_{\text{FOAT}} \to 0.
\end{equation}
Consequently, we obtain:
\begin{equation}
  \lim_{t\to\infty} \|\tilde{g}_t\|^2 = 0,
\end{equation}
which confirms that \( \Theta_t \) converges to a stationary point of the global objective \( F(\Theta) \).

\subsection{Conclusion}

We have established the convergence of \our, a federated fine-tuning framework integrating DLCT, GPO, and FOAT. Under standard smoothness and bounded variance assumptions, \our ensures a monotonically decreasing loss with a well-chosen learning rate \( \eta_t \) and properly tuned hyperparameters ($\lambda$ and $T$). This guarantees convergence to a stationary point, validating its theoretical soundness and practical efficiency in resource-constrained federated settings.

\section{More Experimental Details}\label{appendx_implementation_details}

\subsection{Evaluation Metrics}

We adopt task-specific evaluation metrics to assess model performance across both text classification and instruction tuning tasks~\cite{xu2025videoeraser,xu2024copyrightmeter,xu2026bridging}. For text classification on YELP-P, AGNEWS, YAHOO, and 20NEWS, we report the accuracy, which reflects the proportion of correctly predicted labels and enables consistent comparison across datasets with varying label granularities. For instruction tuning, we follow the official evaluation protocols defined by each benchmark.
Specifically, for MMLU~\cite{hendrycks2020measuring}, we compute the average accuracy over 57 sub-tasks to assess general knowledge and language understanding.
For DROP~\cite{Dua2019DROP}, which targets numerical and discrete reasoning, we report the F1 score to capture partial answer overlap. For BBH~\cite{bbh}, a collection of few-shot reasoning benchmarks, task accuracy is used to evaluate generalization under limited supervision.
For CRASS~\cite{frohberg2022crass}, we adopt counterfactual accuracy to measure robustness under causal perturbations.
All reported results are averaged over five independent runs with different random seeds to ensure statistical reliability and reproducibility~\cite{wang2025indoor,wang2023fedins2}.

\subsection{Experimental Testbed}
Our experiments are conducted on an on-device FL system composed of heterogeneous hardware~\cite{tam2024fedhybrid,tian2024breaking,zhan2024heterogeneity}. These devices possess diverse memory and computational capacities, faithfully emulating real-world deployment conditions. All implementations are built using PyTorch and the Hugging Face Transformers library~\cite{wolf2019huggingface,xu2026agents,xu2026fraudshield,xu2026ididtherelarge,wu2026tsembed}.

\subsection{Implementation Details}

For \our, we set the following hyperparameters unless specified otherwise. The GPO loss balancing weight $\lambda$ is set to 0.1 for BERT and 0.2 for other models. The FOAT threshold $T$ is set to 0.9 for BERT and 0.8 for others. The DLCT co-tuning window size, $Q$, is determined based on the number of co-tuning adapters that the device with the minimum memory capacity can support.

\subsection{Task-Specific Setting}

The training setting is tailored to each task:
\begin{itemize}
\item \textbf{For Text Classification:} In each training round, 15 devices are randomly selected based on memory capacity. Each device performs one local epoch using the SGD optimizer with a batch size of 8. The input sequence length is set to 64 tokens for AGNEWS and 256 tokens for other datasets~\cite{wudevelopmental}.
\item \textbf{For Instruction Tuning:} In each round, 10\% of devices are randomly selected. Each device performs 10 local training steps using the AdamW optimizer~\cite{loshchilov2017decoupled} with a batch size of 16. The maximum sequence length is set to 512 tokens.
\end{itemize}
To ensure a fair comparison, the federated fine-tuning process for all baseline methods is continued until the global model converges.

\section{Baseline Description}\label{appendix_baseline}

\noindent\textbf{Memory-Unaware Methods:} 1) Linear Probing~\cite{kornblith2019better}: This method only fine-tunes the output layer while keeping the rest of the model frozen. 2) FedAdapter~\cite{cai2022fedadapter}: This method dynamically configures adapter modules to accelerate model convergence. 3) C2A~\cite{kim2023client}: This approach leverages a hypernetwork to generate device-specific adapters, thereby addressing data heterogeneity across devices.
    
\noindent\textbf{Memory-Aware Methods:} 4) FwdLLM~\cite{xu2023fwdllm}: A backpropagation-free method that leverages zeroth-order optimization to fine-tune the global model, thereby eliminating the need to store intermediate activations during training.
5) FedKSeed~\cite{qin2023federated}: This method employs zeroth-order optimization with a finite set of random seeds to perform full-parameter fine-tuning. 6) FLoRA~\cite{wang2024flora}: This method addresses memory constraints by assigning heterogeneous LoRA ranks across devices based on their available memory capacity. 7) FedRA~\cite{su2024fedra}: This method addresses memory constraints by randomly assigning model update tasks to participating devices based on their memory capacities.

\section{More Experimental Results}

\definecolor{Maroon}{HTML}{AE3135}
\definecolor{BLUE}{HTML}{6466AE}
\definecolor{my-green}{HTML}{8ECFC9}
\definecolor{my-yellow}{HTML}{FFBE7A}
\definecolor{my-blue}{HTML}{82B0D2}

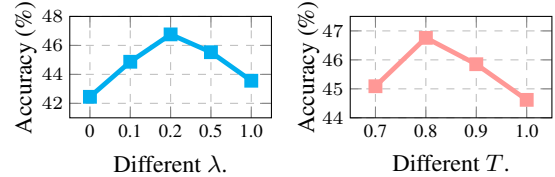
\begin{figure}[!t]
\centering
\pgfplotsset{width=0.55\linewidth,height=0.38\linewidth,compat=1.15}
% \tiny
\begin{tikzpicture}
\centering
\scriptsize{
\begin{axis}[
    at={(0em,0em)},
    xlabel={Different $\lambda$.},
    ylabel={Accuracy (\%)},
    xmin=0.05, xmax=0.55,
    ymin=41, ymax=48,
    xtick={0.1, 0.2, 0.3, 0.4, 0.5},
    ytick={42, 44, 46, 48},
%    legend pos=south east,
    ymajorgrids=true,
    xmajorgrids=true,
    grid style=dashed,
    xticklabels={0, 0.1, 0.2, 0.5, 1.0},
    x label style={at={(axis description cs:0.5,-0.3)},anchor=north, font=\small},
    y label style={at={(axis description cs:-0.13,0.5)},anchor=south, font=\small},
    legend style={
    	at={(0.5,0.1)},
    	anchor=south,
    	legend columns=1,
    	nodes={scale=0.8, transform shape}}
]
\addplot[
    color=cyan,
    mark=square*,
    mark size=1.5pt,thick,line width=1.8pt,
    mark options={fill=cyan,draw=cyan,line width=2.2pt}
    ]
    coordinates {
    (0.1, 42.45)
    (0.2, 44.87)
    (0.3, 46.76)
    (0.4, 45.53)
    (0.5, 43.56)
    };
\end{axis}

\begin{axis}[
	at={(15em,0em)},
    xlabel={Different $T$.},
    ylabel={Accuracy (\%)},
    xmin=0.05, xmax=0.45,
    ymin=44, ymax=47.5,
    xtick={0.1, 0.2, 0.3, 0.4},
    ytick={44,45,46,47},
%    legend pos=south east,
    ymajorgrids=true,
    xmajorgrids=true,
    grid style=dashed,
    xticklabels={0.7, 0.8, 0.9, 1.0},
    x label style={at={(axis description cs:0.5,-0.3)},anchor=north, font=\small},
    y label style={at={(axis description cs:-0.12,0.5)},anchor=south, font=\small},
    legend style={
    	at={(0.43,0.1)},
    	anchor=south,
    	legend columns=1,
    	nodes={scale=0.9, transform shape}}
]
\addplot[
    color=myred,
    mark=square*,
    mark size=1.5pt,thick,line width=1.8pt,
    mark options={fill=myred,draw=myred,line width=2.2pt}
    ]
    coordinates {
    (0.1, 45.09)
    (0.2, 46.76)
    (0.3, 45.85)
    (0.4, 44.62)
    };

\end{axis}
}
\end{tikzpicture}
\vspace{-2mm}
% \caption{Sensitivity Analysis on MMLU (LLaMA2-7B).}
\caption{Sensitivity analysis of key hyperparameters on the MMLU evaluated with the LLaMA2-7B.}
\vspace{-4mm}
\label{fig_sensitivity}
\end{figure}

\subsection{Sensitivity Analysis}

We conduct a sensitivity analysis on our key hyperparameters, $\lambda$ and $T$, to demonstrate the robustness of \our. The analysis, performed on the LLaMA2-7B model (Figure~\ref{fig_sensitivity}), shows that optimal performance is achieved with $\lambda=0.2$ and $T=0.8$. Crucially, these optimal values are consistent with our findings on models in the text classification tasks. This indicates that \our is robust to hyperparameter settings across different models and tasks, requiring minimal tuning.

\begin{figure*}[!t]
  \centering
  \includegraphics[width=0.8\linewidth]{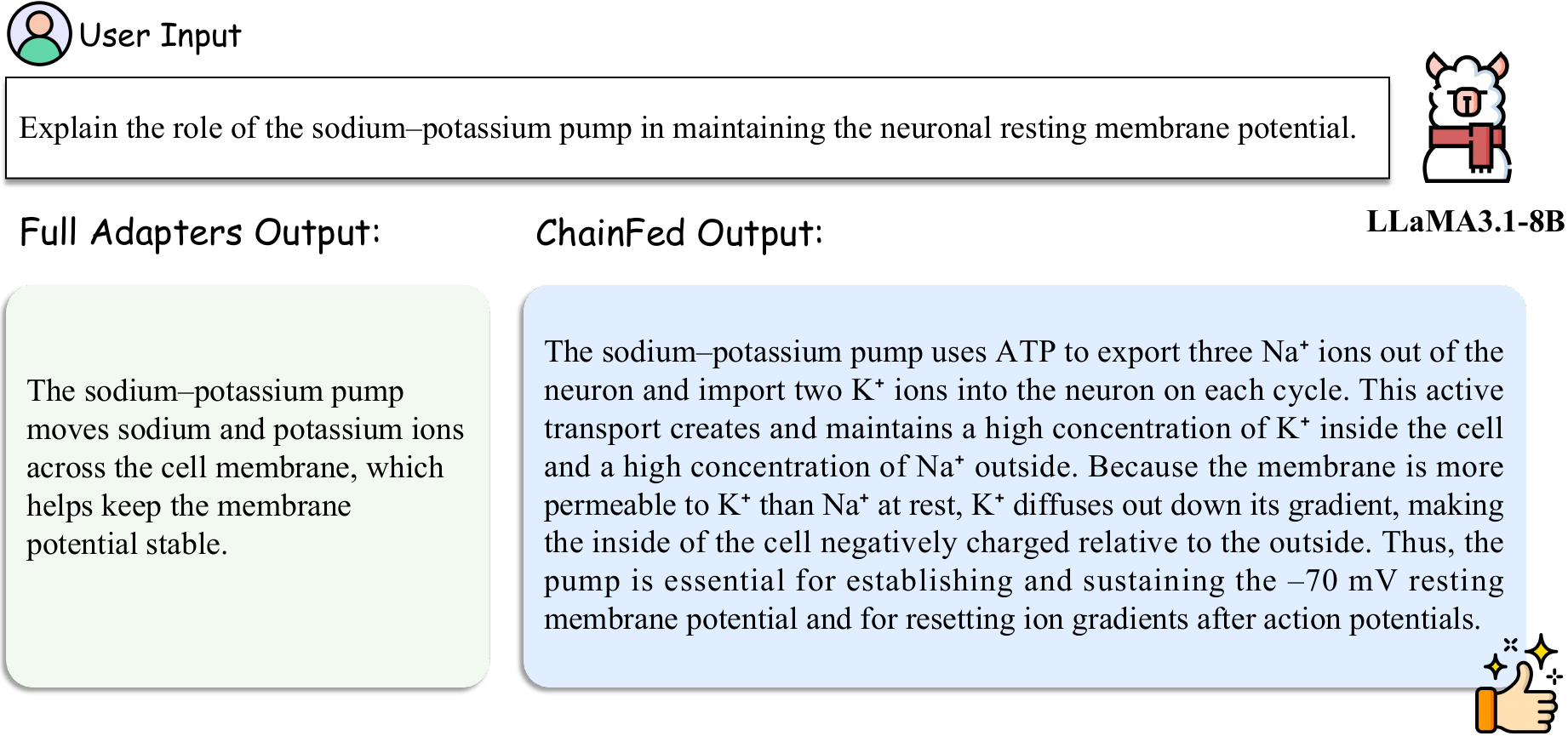}
  \caption{Case Study for LLaMA3.1-8B.}
  \label{fig_case_study}
\end{figure*}

\subsection{Model Generalization}

To verify the generalization capability of \our across different model architectures, we conduct additional experiments on TinyLLaMA and LLaMA3.2-3B, with performance evaluated on the MMLU benchmark. The results show that \our continues to outperform the Full Adapters$^\dagger$. Specifically, with a co-tuning window of $Q=8$, \our surpasses the upper bound by 1.66\% on TinyLLaMA and by a more substantial 4.73\% on LLaMA3.2-3B.
The ability of \our to consistently deliver superior performance across a diverse set of model architectures and scales (from BERT to LLaMA) validates it as a general and robust framework for federated fine-tuning.

\subsection{Case Study: Instruction Tuning Quality}

To provide a qualitative illustration of \our's advantages, we conduct a case study comparing its generated responses to those from the Full Adapters$^\dagger$ for the same instruction. As shown in Figure~\ref{fig_case_study}, \our's responses are demonstrably superior across several axes.
In terms of fluency, \our's output is more coherent, grammatically natural, and flows more smoothly. For contextual alignment, it demonstrates a deeper comprehension of the user's intent, producing answers that are more precisely tailored to the prompt. Finally, regarding informational content, \our provides more detailed and nuanced explanations, covering a broader range of relevant details.

These qualitative improvements suggest that \our's selective, chain-based optimization does more than just save memory; by preserving the foundational knowledge in lower layers, it also leads to better generalization and higher-quality language generation. This highlights the practical effectiveness of \our, demonstrating its ability to produce superior results even while operating under strict memory constraints.

\section{Feasibility and Overhead Analysis}
\label{sec:feasibility_analysis}

A natural concern regarding the chain optimization paradigm is whether the frequent loading and offloading of model layers introduces significant I/O latency, thereby negating the benefits of memory reduction. Additionally, the computational overhead of adaptive mechanisms like FOAT warrants scrutiny. In this section, we analyze these factors to demonstrate that \our maintains high runtime efficiency.

\subsection{I/O Latency Analysis}
The overhead of swapping layers between storage (disk) and memory is negligible compared to the computational cost of gradient calculation.

\begin{itemize}
    \item \textbf{Magnitude Disparity:} On modern hardware, loading a transformer layer takes milliseconds, whereas the forward and backward propagation for that layer takes seconds.
    \item \textbf{Unified Memory Architecture:} On edge devices (e.g., Jetson Orin, mobile phones) which are the target deployment environment for \textsc{ChainFed}, the Unified Memory Architecture (UMA) allows the CPU and GPU/NPU to share memory pointers. This eliminates the need for costly data copying between host and device memory during the offloading process, further minimizing write-back latency.
\end{itemize}

\subsection{Asynchronous Pipelining}
To further mitigate I/O costs, \textsc{ChainFed} employs an asynchronous ``Compute-Prefetch-Evict'' pipeline that hides I/O latency behind computation.

\begin{itemize}
    \item \textbf{Mechanism:} While the device performs the computation for the current active layer $L_i$, the system asynchronously writes the parameters of the previous layer $L_{i-1}$ back to storage and prefetches the parameters of the next layer $L_{i+1}$ into a reserved memory buffer.
    
    \item \textbf{Result:} Since the computation time ($T_{comp}$) significantly exceeds the I/O time ($T_{I/O}$), the read/write operations complete before the computation finishes. Thus, the latency is determined by $T_{comp}$, rendering the I/O overhead invisible in the critical path of execution. Consequently, \textsc{ChainFed} incurs almost zero additional latency attributed to I/O. 
    
    % Furthermore, given that \textsc{ChainFed} only updates a subset of layers in each round, the required computational load is inherently low, resulting in high overall training efficiency.
\end{itemize}

\subsection{Efficiency and Feasibility of FOAT}
\label{sec:foat_analysis}

% A potential concern is that calculating layer similarity for FOAT might impose heavy computational burdens or require memory capacities beyond the limits of edge devices. We address these concerns by clarifying the efficiency and feasibility of our design:
A natural concern is whether computing layer similarity for FOAT imposes prohibitive computational or memory overheads on edge devices. We address this by detailing the efficiency and practical feasibility of our design:

\begin{itemize}
    \item \textbf{Computational Efficiency:} 
    First, the CKA computation is a one-time pre-training setup step (Phase 1 in Algorithm~\ref{alg:profit}). It does not recur during the federated rounds. Furthermore, to minimize overhead, we perform this profiling using only a single mini-batch of local data. Consequently, the cost is negligible compared to the full training process.

    \item \textbf{Resource Feasibility:} To ensure feasibility on devices with limited memory (where loading the full model for profiling is impossible), we implement a block-wise inference strategy. This protocol executes in three phases:
    \begin{enumerate}
        \item \textit{Budget-Aware Partitioning:} The model is dynamically divided into manageable blocks. The device loads only the specific block of layers that satisfies its current available memory constraints.
        \item \textit{Transient Computation with Immediate Eviction:} Upon performing the forward pass on the current block, the system caches the resulting intermediate activations (feature maps) and immediately evicts the layer parameters from memory to release resources.
        \item \textit{Recursive Activation Passing:} The subsequent block of layers is then loaded, taking the cached activations from the previous step as input.
    \end{enumerate}
    By decoupling the inference process from total model size, this strategy ensures that peak memory usage remains strictly bounded by the size of a single block, enabling large-model profiling on constrained devices.
\end{itemize}

% \section{Discussion and Additional Analysis}
\section{Further Analysis}
\label{appendix_discussion}

% In this section, we provide a deeper analysis of \our's design principles, theoretical foundations, and practical implications. We explore key aspects including performance characteristics, computational efficiency, robustness properties, and scalability considerations to offer comprehensive insights into the method's behavior and potential applications.

\subsection{Performance Advantages of \our}

Empirically, \our outperforms the idealized Full Adapters$^{\dagger}$ baseline across all evaluation settings, achieving average gains of 2.86\%, 2.46\%, and 1.61\% on DistilBERT, BERT, and RoBERTa, respectively. This performance advantage stems from two key mechanisms:

\begin{itemize}
    \item 1) \textbf{Selective Layer Adaptation.} By leveraging FOAT to identify and fine-tune only task-critical layers, \our preserves valuable pre-trained knowledge in lower layers that encode general-purpose linguistic patterns. In contrast, updating all layers indiscriminately can disrupt these foundational representations. 
    % Our experiments further validate this: freezing lower layers ($T=0.8$) consistently outperforms full-chain tuning ($T=1.0$), with improvements of up to 4.43\% on AGNEWS.
    
    \item 2) \textbf{Enhanced Generalization.} The combination of DLCT and GPO promotes more stable and generalizable feature learning. DLCT ensures semantic coherence across layers, while GPO prevents over-specialization by maintaining global awareness. This synergistic effect leads to better generalization compared to end-to-end training, which may overfit to local patterns in federated settings.

\end{itemize}

% These results demonstrate that these techniques transform chain optimization from a memory-saving strategy into a performance-enhancing paradigm.

\subsection{Efficiency Analysis}

% Beyond memory reduction, \our also achieves computational and communication efficiency. 
In addition to significantly reducing the memory footprint, \our delivers substantial gains in both computational and communication efficiency.

\begin{itemize}
    \item \textbf{Computational Efficiency.} Although adapters are trained sequentially, the total number of forward and backward passes remains comparable to end-to-end training. Moreover, selectively fine-tuning further accelerates model convergence. Empirically, \our achieves 2.02$\times$ faster convergence than Full Adapters$^{\dagger}$ on AGNEWS.

    % \item \textbf{Computational Efficiency.} The chain optimization paradigm introduces minimal computational overhead. Although adapters are trained sequentially, the total number of forward and backward passes remains comparable to end-to-end training. The overhead from DLCT (co-tuning $Q$ adapters) is offset by skipping training on frozen lower layers via FOAT. Empirically, \our achieves 2.02$\times$ faster convergence than Full Adapters$^{\dagger}$ on AGNEWS.

    \item \textbf{Communication Efficiency.} Communication costs are significantly reduced through two mechanisms: (1) FOAT reduces the number of layers requiring updates, directly decreasing communication volume; (2) the sequential nature ensures that in each round, only adapters within the co-tuning window need synchronization. Our results show that communication can be reduced by up to 3.47$\times$ compared to Full Adapters$^{\dagger}$.

\end{itemize}

\section{The Use of Large Language Models}

This manuscript utilized LLMs strictly for the purpose of language editing and textual polishing to enhance presentation quality. We declare that the novel ideas, methodological framework, experimental execution, and data analysis are the original work of the authors. All content modified by AI tools has been carefully reviewed and validated by the authors to ensure accuracy.

\end{document}